\documentclass[journal]{IEEEtran}
\usepackage{amsmath,amsfonts,amssymb}
\usepackage{algorithmic}
\usepackage{algorithm}
\usepackage{array}
\usepackage[caption=false,font=normalsize,labelfont=sf,textfont=sf]{subfig}
\usepackage{textcomp}
\usepackage{stfloats}
\usepackage{url}
\usepackage{verbatim}
\usepackage{graphicx}
\usepackage{cite}
\usepackage{xcolor}
\usepackage{graphicx}
\usepackage{textcomp}
\usepackage{mathtools}
\usepackage{hhline}
\usepackage{bm}
\usepackage{cleveref}
\usepackage{stackengine}
\usepackage{stmaryrd}
\usepackage{booktabs}
\usepackage{caption} 
\usepackage{dirtytalk}
\definecolor{purple}{cmyk}{0,0.98,0.41,0.3}
\definecolor{myyellow}{cmyk}{0,.15,1,0}

\DeclareMathOperator*{\argmin}{arg\,min}
\DeclareMathOperator{\sem}{\Phi}
\DeclareMathOperator{\LS}{\mathcal{L}}
\DeclareMathOperator{\bx}{\mathbf{x}}
\DeclareMathOperator{\bz}{\mathbf{z}}

\DeclareMathOperator{\dbi}{\mathcal{I}_{\mathcal{X}}}
\DeclareMathOperator{\dico}{\mathbf{T}}
\DeclareMathOperator{\dicoq}{\tilde{\mathbf{T}}}

\begin{document}

\title{SMIC: Semantic Multi-Item Compression based on CLIP dictionary}

\author{\IEEEauthorblockN{Tom Bachard, Thomas Maugey} \\
    \IEEEauthorblockA{IRISA, INRIA, Univ Rennes}
    
\thanks{This work was funded by the French National Research Agency (\emph{MADARE}, Project-ANR-21-CE48-0002 and \emph{Contrats doctoraux en intelligence artificielle}, ANR-20-THIA-0018).}

}

% The paper headers
\markboth{Journal of \LaTeX\ Class Files,~Vol.~14, No.~8, August~2021}%
{Shell \MakeLowercase{\textit{et al.}}: A Sample Article Using IEEEtran.cls for IEEE Journals}

%\IEEEpubid{0000--0000/00\$00.00~\copyright~2021 IEEE}
% Remember, if you use this you must call \IEEEpubidadjcol in the second
% column for its text to clear the IEEEpubid mark.

\maketitle

\begin{abstract}
Semantic compression, a compression scheme where the distortion metric, typically MSE, is replaced with semantic fidelity metrics, tends to become more and more popular. Most recent semantic compression schemes rely on the foundation model CLIP. In this work, we extend such a scheme to image collection compression, where inter-item redundancy is taken into account during the coding phase. For that purpose, we first show that CLIP's latent space allows for easy semantic additions and subtractions. From this property, we define a dictionary-based multi-item codec that outperforms state-of-the-art generative codec in terms of compression rate, around $10^{-5}$ BPP per image, while not sacrificing semantic fidelity. We also show that the learned dictionary is of a semantic nature and works as a semantic projector for the semantic content of images.
\end{abstract}

\begin{IEEEkeywords}
Compression, Semantics, Multi-item, Deep-learning
\end{IEEEkeywords}

\section{Introduction}

Since decades, strong research efforts have been spent to improve image compression regarding the rate-distortion performance. However, even with impressive improvements \cite{bross2021overview, girod1997performance, richardson2011h, vanne2012comparative}, efforts have to be made to deal with the enormous amount of data transmitted every day \cite{webdomo}. Such a way to cope with the never-ending increasing quantity of data is to change the paradigm away from the classical rate-distortion evaluation.

In most image collections (private or public), there exist some redundancies that are rarely considered during storage. These statistics could, however, lead to more efficient compression. Exploiting such inter-image redundancy during coding is the goal of the so-called multi-item compression (MIC) paradigm. The nature of the redundancies has a great impact on the types of techniques used by the compression scheme. Most of the existing algorithms, \cite{MIC1, MIC2, MIC3, MIC4, MIC5, MIC6, MIC7}, have tracked the redundancy residing at the pixel level (as in the classical image/video compression paradigm). However, data collection often presents more complex relationships between the images. Indeed, even with pixel-wise different content, images may describe the same scene or object. In that case, we talk about \textit{semantic redundancy}. In previous work, \cite{bachard2022}, 
we have shown that exploiting such redundancy in a conventional MSE-based compression framework is possible, but may come with some performance loss.

Recently, Semantic Compression (SC) architecture has been raised to explore extremely low bitrate. These architectures rely on
\cite{blau2017}, where it is shown that, at extremely low compression bitrates, one has to choose between perception -- the quality of the outputs -- and distortion -- to what extent the outputs are far from the inputs. Semantic compression thus leaves the pixel fidelity criterion, typically measured with MSE, PSNR or SSIM, and replace it with a semantic fidelity distance. Indeed, the crucial difference between conventional compression and SC is the fact that in the latter, the images are generated, instead of reconstructed, from a high-level description of the inputs. The motivation for such a framework is that the important information in an image does not lie at the pixel level, but rather a higher, more semantic, level. Such frameworks are typically used for cold data \cite{chaudhuri2018decision} or with the coding for machines paradigm \cite{codingformachine}. In the SC framework, an encoder, \emph{e.g.} \cite{llm} or \cite{sc_net}, represents the input semantics in a compact form, while an image generator, \emph{e.g.,} \cite{goodfellow2020generative} or \cite{ho2020}, synthesize an image sharing the same high-level description as the input. In \cite{bachard2024}, we showed that CLIP, together with UnCLIP, proposes a suitable semantic description for compression. 
However, these semantic compression techniques have never been extended to the simultaneous compression of multiple images.

In this work, we propose to explore multi-item compression in the context of the semantic compression paradigm. First, we formally define multi-item compression (MIC) and semantic compression (SC) and how we propose to fuse them into semantic multi-item compression (SMIC). In a second section, we define the methodology, mainly inherited from MIC and SC, used in this paper, especially the definition of CLIP. In the following section, we define, prove, and propose limitations to semantic linear operations inside CLIP's latent space. Section V proposes to use the previous property to learn how to create a dictionary from a database and how to project and recreate the latent vectors from this dictionary. Next, we studied the semantic properties of this dictionary: semantic conservation and semantic separation. The last section derives a multi-item generative compression scheme from the aforementioned learned semantic dictionary.  In this last section, the proposed SMIC framework is compared to state-of-the-art single item compression (SIC) schemes and is shown to have attains better compression rates while maintaining a comparative semantic fidelity.

The main contributions of this work are the following:
\begin{itemize}
    \item We define and prove that CLIP's latent space additions and subtractions, up to renormalization, induce semantic additions and subtractions in the images are generated with UnCLIP;
    \item We demonstrated that creating a dictionary with CLIP's latent vectors from a data collection is possible, and that this dictionary is of a semantic nature: the atoms represent high-level concepts. Moreover, semantic separation of concepts is possible regarding the generation with the projections or with the residuals;
    \item We proposed a semantic multi-item compression pipeline based on CLIP and on semantic dictionary that outperforms the state-of-the-art single item compression algorithms at extremely low bitrates (around $10^{-5}$ BPP) while conserving semantic fidelity.
\end{itemize} 

\section{Problem formulation}
\label{sec:MIC}

    This work proposes a solution to compress a large collection of images using semantic compression techniques. In this section, we first formulate the general multi-item compression problem. In a second part, we introduce the semantic compression framework. Finally, we formulate the problem of \emph{semantic multi-item compression (SMIC)} tackled in this work, which is the combination of the two aforementioned frameworks.

    \subsection{Multi-item Compression}

    Multi-item compression (\emph{MIC}) is a coding framework that aims at compressing a collection of images $\mathcal{X} = (\bx_1, \dots, \bx_N)$ exploiting the inter-item redundancy. The efficiency of such a coding scheme is measured both in terms of compression rate and reconstruction error.
    
    \begin{figure}[htbp]
        \centering
        \hspace*{-.3cm}
        \includegraphics[width=.5\textwidth]{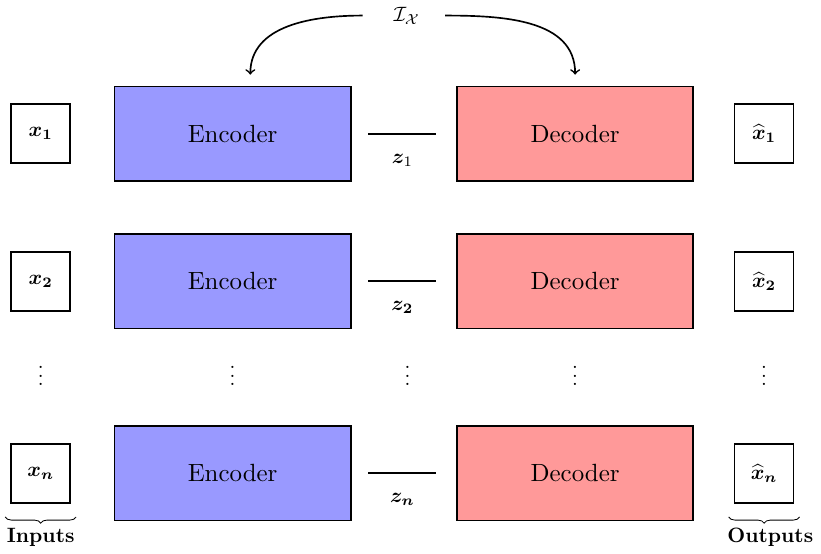}
        \caption{Multi-item compression with individual image coding. $\dbi$ describes the database's statistics used for individual encoding and decoding.}
        \label{fig:mic}
    \end{figure}

    The MIC scheme is given in Fig. \ref{fig:mic}. Each image $\bx_i$ of a data collection $\mathcal{X}$ is transformed into a bit stream $\bz_i$ via an encoder. This encoder takes the $\mathcal{X}$'s statistical model, $\dbi$, as side information. The bit-stream $\bz_i$ is then decoded as an image $\hat{\bx}_i$ at the user side with a decoder using the same side information $\dbi$.  
    
    In the described MIC framework,  each item is encoded individually so that new images can be added to the database without having to re-compress every other images again. Moreover, the extraction and encoding of the database statistics $\dbi$ is only done once with the original database, and future added images are supposed to be correlated with this original database. The compression rate $R$ for the whole database is then the length of the bit stream $R=\sum_{i=1}^{N}\mathcal{R}(\bz_i)$, to which we add the weight of the dataset statistics $\mathcal{R}(\dbi)$ used by the codec. 

    All in all, the multi-item compression problem can be stated as the following minimization problem, where $d$ is a distortion metric between the original images and the generated ones and $\tau$ a threshold ensuring a maximal acceptable error between them:
    \begin{gather}
        \min \sum_{i=1}^N \mathcal{R}(\bz_i) + \mathcal{R}(\dbi) \ \text{  s. t. } \\
        \forall i\in \llbracket 1, N \rrbracket,\ d_\Phi\left(\bx_i,\ \hat{\bx}_i\right ) < \tau .\nonumber
    \end{gather}

    To reduce the rate of coded images, one needs to find redundancies between the $\bz_i$. In \cite{bachard2022}, we have shown that, if the correlation resides at the semantic level (and not at the pixel level), one must adapt the latent space so that it also captures semantic information. This has, however, a cost in terms of compression efficiency. This is why we decided to study the multi-item compression problem in the context of semantic compression, as defined in the next two sections.

    \subsection{Semantic Compression}

    \begin{figure}[htbp]
        \centering
        \includegraphics[width=0.5\textwidth]{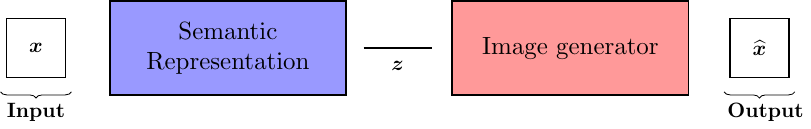}
        \caption{Generative compression. The generated images are evaluated in terms of semantic fidelity and visual quality.}
        \label{fig:framework}
    \end{figure}

    The work in \cite{blau2017} showed that at very low bitrates, there exists a trade-off between distortion and perception for MSE-based compression schemes. To overcome this limitation, one can choose to replace the pixel-wise metric with semantic-based metrics. As a consequence, the compressed description of the image only captures the high-level information about the image. This information is used to guide and control an image generator acting as a decoder. This framework is called \emph{semantic compression (SC)}.
    
    Fig.~\ref{fig:framework} presents the semantic based generative compression framework. The input image $\bx$ is encoded into a latent semantic representation $\bz$ via a semantic encoder. An image generator, acting as a decoder, reconstructs the decoded input $\mathbf{\tilde{x}}$ using the semantic present in the latent representation. Unlike classical compression, the reconstruction error is not evaluated with a classical pixel-based loss (MSE), but rather to what extent the semantic of the generated images is close to the semantic of the inputs. Let us assume that the semantic information of an image $\bx$ is given by a function $\Phi$. We can then write the semantic distance between an input and the generated image $\mathbf{\tilde{x}}$ as $d_\Phi(\bx, \mathbf{\tilde{x}})=d(\Phi(\bx), \Phi(\mathbf{\tilde{x}}))$. However, this semantic distance does not guarantee a visually pleasant image. So, to evaluate the perceptual quality of the image, we use a no-reference metric $\Psi$.
    
    Given $\Psi$ and $\Phi$, we define the problem as minimizing the bitrate under semantic coherence $\tau_\Phi$ and realism $\tau_\Psi$ constraints: 
    
    \begin{gather}
        \label{eq:framework}
        \min \mathcal{R}(\bz) \text{  s.t. } \\
        \Psi(\bx) > \tau_{\Psi}  \text{ and } d_\Phi(\bx, \mathbf{\tilde{x}})<\tau_{\Phi} \nonumber
    \end{gather}

    Because of the $d_{\Phi}$ distance, semantic coding schemes lead to compressed descriptions representing the semantic information about the image. They enable, generally, to explore ultra-low bitrate. For our scenario where a data collection has correlations at the semantic level, this property is interesting as this correlation can be directly reflected in the compressed description. That is the reason we decided to mix semantic compression with multi-item compression in the next section. 

    In this work, the semantic representation function is CLIP \cite{clip}. Specifically, we use the Vital/14 version of the model. In this version, images are encoded in a $768$-dimensional vector coded on $16$-bits. For the image generator, we use the Stable unCLIP \cite{unclip} model, a CLIP fine-tuned latent diffusion model based on the Stable Diffusion model \cite{stablediffusion}. The used weights can be found here\footnote{https://huggingface.co/docs/diffusers/api/pipelines/stable\_unclip}. We specify that CLIP and Stable unCLIP are \emph{not} fine-tuned nor retrained for any of the experiments presented in this work.
    
    \subsection{Semantic Multi-Item Compression}

\begin{figure}[htbp]
        \centering
        \includegraphics[width=.5\textwidth]{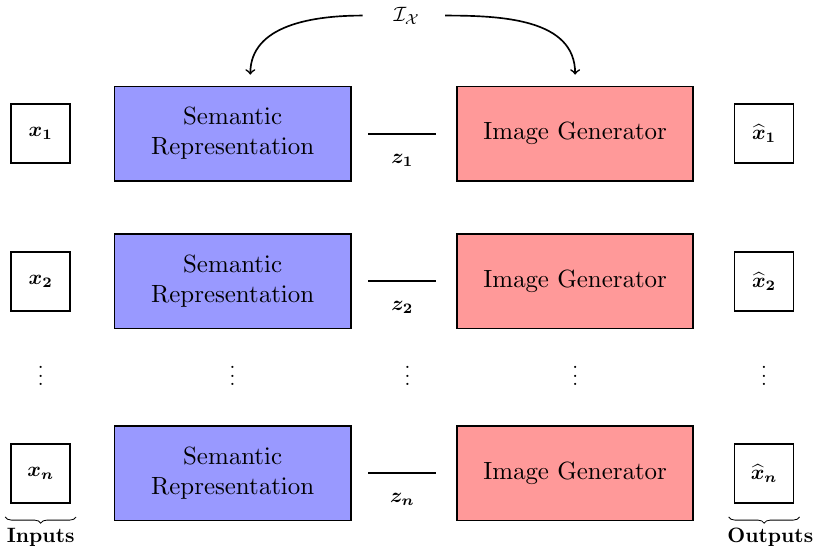}
        \caption{Semantic Multi-item compression. $\dbi$ describes the database's statistics used for individual encoding and decoding.}
        \label{fig:migc}
    \end{figure}

    The framework developed in this work will take advantages of both the previously introduced frameworks. \emph{Semantic Multi-item compression (SMIC)}  is a compression scheme that aims at compressing a database of images at a very low bitrate by taking advantages of the redundant semantic present in the database. %To do so, we are going to adapt the MIC framework with the SC framework. 
    SMIC framework is presented in Fig.~\ref{fig:migc}.

    For $\mathcal{X}$, a given collection of $N$ images, SMIC framework aims at minimizing the following problem, where $\tau_{\Phi}$ and $\tau_{\Psi}$ are respectively the semantic coherence threshold and the realism threshold:

    \begin{gather}
        \min \sum_{i=1}^{N} \mathcal{R}(\bz_i) + \mathcal{R}(\dbi) \text{\ \ \ s.t.} \\
        \forall i\in \llbracket 1, N \rrbracket,\ \Psi(\hat{\bx}_i) > \tau_{\Psi} \text{\ \  and\ \ } d_{\Phi}(\bx_i,\hat{\bx}_i) < \tau_{\Phi}  \nonumber
    \end{gather}

    Solving the SMIC problem is to be able to capture the redundancy between the $\mathbf{z}_i$ with a model $\mathcal{I}_{\mathcal{X}}$, and to exploit this redundancy to reduce the cost of describing each $\mathbf{z}_i$. We present in Section~\ref{sec:operation} some properties of CLIP latent space that can be useful for solving this challenge.

\section{Semantic linearity in CLIP's latent space}\label{sec:operation}

    Image semantics means the high-level information depicted by an image. More specifically, each pixel value only gives a pointwise color information, and the concatenation of these pixels forms more general concepts such as contours, textures, shapes, etc. Going further, the concatenation of these concepts leads to a high-level interpretation of the scene described by the image (\textit{e.g.,} objects, actions, atmosphere, feelings). These elements are typically referred to as the general concept of \emph{semantic}. In the following, we denote by $\sem(\bx)$ the semantics of an image $\bx$. 
    
    In practice, extracting the image semantics is done with complex tools that are highly non-linear (\emph{e.g.,}  CNN, deep models). The model CLIP (of interest in this work) enables to describe the image semantics into a $768$-dimensional vector, that most likely relies on a spherical space, as the semantic similarity between two images is given by their angle \cite{bachard2024}. At a first sight, the shape of CLIP embedding space and the deep model that enables to build the CLIP latent make us think that the CLIP space is highly non-linear. However, in this work, we prove that operations in the semantic world (such as addition/subtraction of two semantic concepts) naturally translate into the CLIP domain  (as a simple addition/subtraction between the CLIP vector).

    More specifically, for two images $\bx_1$ and $\bx_2$, we prove that CLIP, for any real $\lambda$, verifies:
    \begin{gather} \label{eq:semlin}
        \sem(\text{CLIP}(\bx_1) + \lambda\text{CLIP}(\bx_2)) =\\
        \sem(\text{CLIP}(\bx_1)) + \lambda \sem(\text{CLIP}(\bx_2)) \nonumber
    \end{gather}
    In the proposed property, $\lambda$ encapsulates the type of the operation. If $\lambda>0$, we are adding the semantics of the two images. On the other hand, if $\lambda<0,$ we are subtracting the concepts present in the second image from the first one. All in all, $|\lambda|$ controls the magnitude of the operation.
    
\begin{figure*}[htbp]
        \centering
        \includegraphics[width=.95\textwidth]{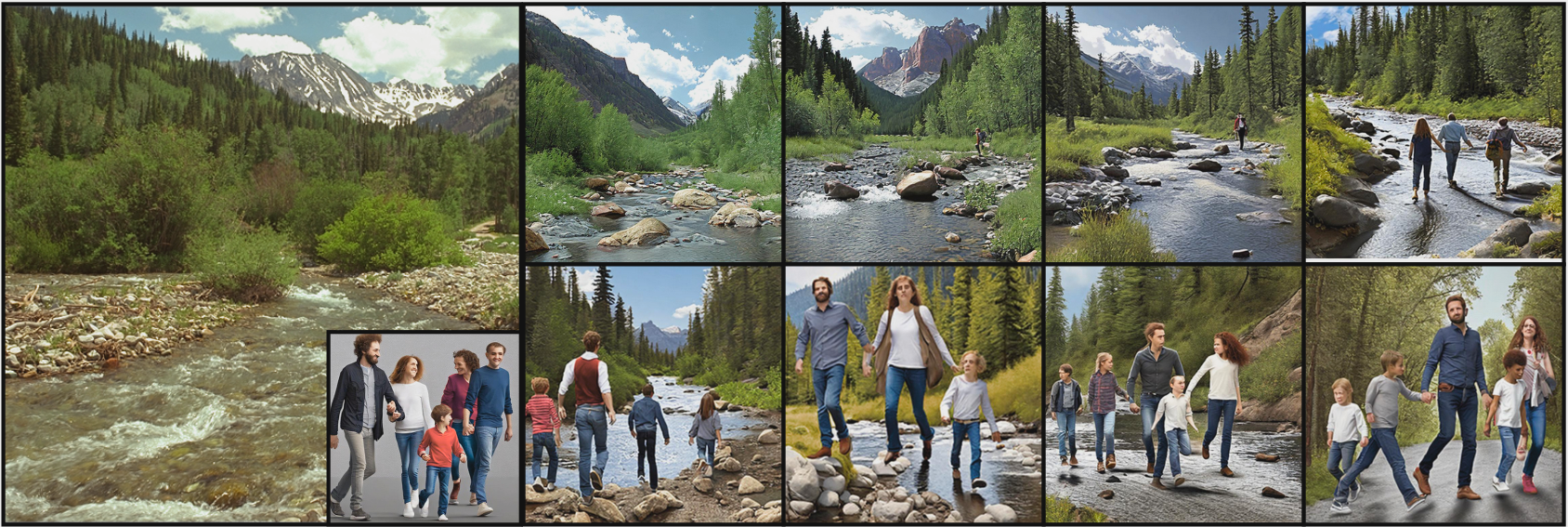}
        \caption{Progressively adding people to the landscape from \cite{kodak} (Left) Input images $\bx_1$ and $\bx_2$. (Right) Top to bottom, left to right: Images generated from $f(\bx_1)+\alpha f(\bx_2)$. Where $\alpha=i/4,\ i\in [1...8]$.}
        \label{fig:add_lin}
    \end{figure*}

    To demonstrate that  property, we visually show that one can add or subtract latent vectors and that such operations induce semantic addition and subtraction in the generated images. As shown in \cite{bachard2024}, we also proceed to a normalization of the resultant latent vector, as unCLIP has been trained to generate images from latent vectors around a given norm ($\sim 20$ in this case). The results of these operations are presented in Fig. \ref{fig:add_lin}, when $\lambda > 0$, and in Fig. \ref{fig:sub_lin}, when $\lambda < 0$ (more results are given in the supplementary materials). In the proposed examples, we observe in both cases, addition  and subtraction, that the operation is progressive. Indeed, the greater $|\lambda|$, the greater the semantic modification. For the addition, we observe that the individuals are getting more and more present in the resulting image, to the point where, for high values of $\lambda$, some of the original semantic is lost in the generative process. Regarding the subtraction process, we observe that, starting from the same input image, we can either delete the \emph{river} part of the original picture or either the \emph{forest} part (see the supplementary materials). In the same vein as the additions example, we observe here that $\lambda$ controls to what extent the concept is deleted in the resulting image. From these examples, we conclude that, indeed, the CLIP-unCLIP proposed generative codec fulfills Eq.~(\ref{eq:semlin}). 
    
    In our experiments, we have observed this linearity of the CLIP latent for many examples. However, there exist some case where this addition does not work (such examples are shown in the supplementary material). This corresponds to cases where the semantics concepts that are added has never been seen together during the CLIP's training. In other words, this linearity property is satisfied for natural combination of semantic concepts, as they could be seen in real images.
    
    \begin{figure*}[htbp]
        \centering
        \includegraphics[width=.95\textwidth]{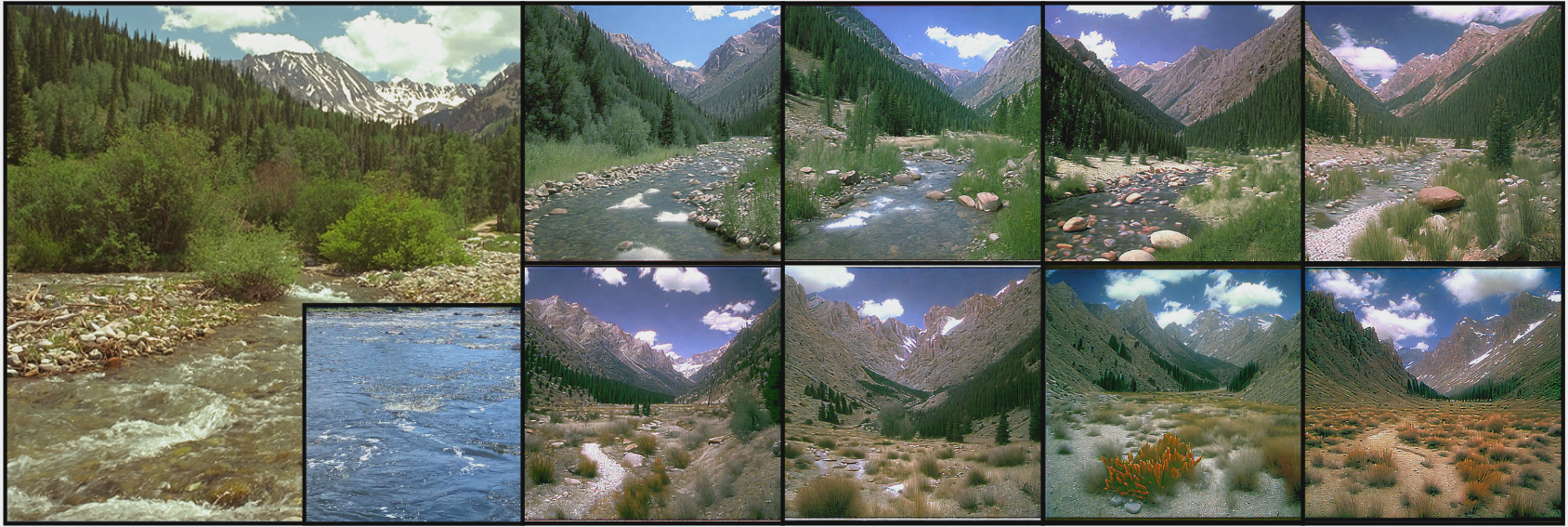}
    
        \vspace{.5cm}
        
        \caption{Progressively removing the river from the landscape from \cite{kodak}. (Left) Input images $\bx_1$ and $\bx_2$. (Right) Left to right: Images generated from $f(\bx_1)-\alpha f(\bx_2)$. Where $\alpha=i/8,\ i\in[1...8]$.}
        \label{fig:sub_lin}
    \end{figure*}

\section{Learning a semantic dictionary from an image database}

    In this section, we show that the semantic property discussed in the previous section can be used to derive a dictionary of atoms encapsulating the image collection's semantics. Furthermore, we show that this dictionary can serve to project the latent vectors into sparse coefficients that can be later used to reconstruct their original latent vector inputs. 
    
    \subsection{Motivation}

    We have shown that additions and subtractions in the latent space of CLIP result in semantic additions and subtractions in the pixel domains when the images are generated. From Eq.~(\ref{eq:semlin}), stated for $2$ latent vectors, we can immediately derive an expression for multiple latent vectors. This means that a semantically complex scene $\bx$ can be described as the weighted sum of the latent representation of its components, expressed as simple semantic. Given $(\bm{t}_i)$ a collection of latent vectors, we can express any image $\bx$ as a linear combination of the $\bm{t}_i$:

    \begin{align} \label{eq:decomposition}
        \text{CLIP}(\bx) = \bz = \sum_{j=1}^{n_a} c_j \bm{t_i}
    \end{align}
    Where $n_a$ is the number of concepts spanned by $(\bm{t}_i)$, and each $(c_i)$ are the “intensities” of each concept. For example, a typical mountain view could be described as the sum of simple concepts such as \say{mountain} + \say{forest} + \say{cloudy sky}.

    In this work, we aim at learning a semantic description of the database we encode, so the $\bm{t}_i$ from the previous equation should encapsulate the database semantic. We define $\dico = (\bm{t}_i)$ as the collection of high-level, yet simple, latent vectors that represent the semantics of the data collection. From this collection, we can then encode an image $\bx$ into the coefficients $\mathbf{c} = (c_i)_{1\leqslant i \leqslant N_T}$ such as proposed in Eq.~(\ref{eq:decomposition}). As the goal of the codec is to attain extremely low bitrates, we are looking for $\dico$ to be sufficiently expressive such that the coefficients of most of the images in the collection are sparse. 
    
    In the following, we consider that both the encoder and the decoder needs $\dico$, either, to encode the image into a vector of coefficients or to reconstruct the latent vector from which the images will be generated. We then have to account for the bitrate of $\dico$ in the compression scheme and ensure that its over cost is absorbed for sufficiently small databases. 

    One of the bottlenecks for this compression scheme now becomes, for every image, the list of coefficients and their compression rate. To compress them, we propose to use classical coding tools such as entropic coding, sparsity over the coefficients, and the trade-off between the number of vectors in $\dico$ and the quality of the reconstruction. These are discussed in Sec.~\ref{sec:IoAS}. But first, we focus on discussing $\dico$ and how to learn it.
    
    \subsection{Semantic latent dictionary}

    The previous discussions showed that we are looking for a collection of simple semantic vectors, $(\mathbf{t}_i)$, in which every image of the input data collection can be decomposed. This leads us to define $\dico$ as a dictionary of semantic atoms. Each encoded image $\bz$ can now be considered a collection of coefficients $\mathbf{c}$, encapsulating its semantic.

\begin{figure*}[htbp]
        \centering
        \includegraphics[width=\textwidth]{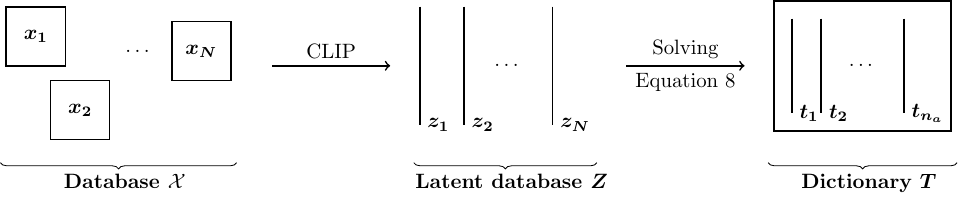}
        \caption{Learning the dictionary from an image collection.}
        \label{fig:learn-dict}
    \end{figure*}

    More formally, we solve the minimization problem presented in Fig.~\ref{fig:learn-dict}. Given a collection of images (that can be the whole image collection $\mathcal{X}$, a sub-part of it, or the collection augmented with any other images), we note $\mathbf{Z}\in\mathbb{R}^{\LS\times N}$ the collection of their latent vectors, such that each column  $\bz_i = \text{CLIP}(\bx_i)$.  The goal of the operation is to find $\mathbf{T}$ a dictionary that covers the space spanned by the database.  To do so, we fix $n_a$, the number of atoms in $\dico$ and we solve the classical dictionary minimization problem:
    \begin{align} 
        \mathbf{T}^{*} = \argmin_{\substack{\mathbf{T}\in\mathbb{R}^{\LS\times n_a}\\\mathbf{C}\in\mathcal{R}^{n_a\times N}}}\frac{1}{2} \| \mathbf{Z} - \dico\mathbf{C} \|^{2}_{2} + \lambda \|\mathbf{C}\|_0 
    \end{align}
    Where $\lambda$ controls the trade-off between the reconstruction error and the sparsity of the coefficients $\mathbf{C}$, representing the column-wise coefficients associated with each latent vector in $\mathbf{Z}$. 
    
    Due to intractability in the solving method, we relax the $\mathcal{L}_0$ parsimony constraint into a $\mathcal{L}_1$ parsimony constraint, and we solve the following problem using gradient descent \cite{scikit}:

    \begin{align} 
        \mathbf{T}^{*} = \argmin_{\substack{\mathbf{T}\in\mathbb{R}^{\LS\times n_a}\\\mathbf{C}\in\mathcal{R}^{n_a\times N}}}\frac{1}{2} \| \mathbf{Z} - \dico\mathbf{C} \|^{2}_{2} + \lambda \|\mathbf{C}\|_1 \label{eq:learn_dict} 
    \end{align}

    \subsection{Interpretation of atom's semantic}
    \label{sec:IoAS}

    \begin{figure*}[htbp]
        \centering
        \includegraphics[width=\textwidth]{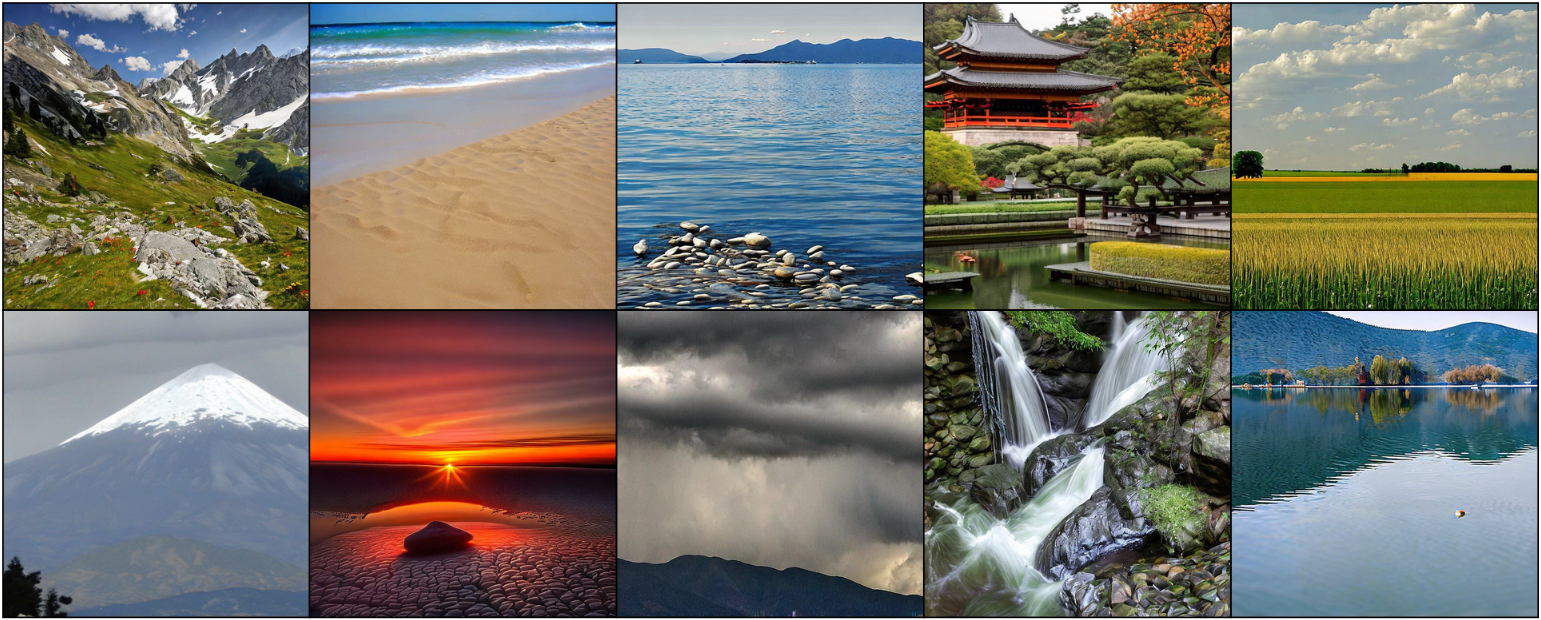}
        \caption{Images generated from the first ten atoms of a dictionary learned of the whole landscape \cite{landscape} dataset.}
        \label{fig:ex_dico}
    \end{figure*}

    Given a dictionary $\dico$ learned using Eq.~(\ref{eq:learn_dict}), we are now interested in the semantic interpretation of each of its atoms $\bm{t}_i$ regarding the original data collection. Fig.~\ref{fig:ex_dico} shows images generated from the first ten atoms of a dictionary of $32$ atoms. We observe that each image, thus each atom, represents simple and unique semantic concepts (\say{mountain}, \say{beach}, \say{sea}, \dots). Similar experiments with different dictionary sizes are proposed in supplementary materials, and they all point towards the same conclusion. The learned dictionaries encapsulate broad but various high-level descriptions of the images present in the database. This property is encouraging for reconstructing images from the atoms, as discussed in the next section.

    Evaluating the reconstruction error in the generative compression framework is different from the pixel-based methods used with classical codecs. Indeed, the output images are produced by an image generator rather than reconstructed by a decoder. As the continuation of \cite{bachard2024}, we evaluate the same semantic coherence metrics: CC \cite{clipscore}, BSS, and CSS \cite{bachard2024}.

    \subsection{Reconstructing images with a semantic dictionary}

    Once the dictionary $\mathbf{T}$ is learned, one can project any latent vector $\bz$ on this basis to obtain its coefficients' representation $\mathbf{c} = [c_1, \dots, c_{n_a}]$ by solving almost the same minimization problem via coordinate descent \cite{lasso_cd}:
    \begin{align} \label{eq:learn_coef}
        \mathbf{c} = \argmin_{\mathbf{c}\in\mathbb{R}^{n_a}}\frac{1}{2} \| \mathbf{z} - \mathbf{T}\mathbf{c} \|^{2}_{2} + \lambda \|\mathbf{c}\|_1
    \end{align}

    Given a semantic dictionary $\dico$ and an image $\bx\in\mathcal{X}$, we are interested in whether the reconstructed latent vector $\bz =\sum_{i=1}^{n_a}c_i\bm{t}_i$ can be used by UnCLIP to generate images $\hat{\bx}$ that are both qualitative and semantically close to $\bx$. The coefficients $(c_i)_{1\leqslant i\leqslant n_a}$ are obtained by solving Eq.~(\ref{eq:learn_coef}).
    
    It has been shown in \cite{bachard2024} that any image generated from a CLIP vector will be qualitative according to several no-reference metrics $\Psi$, as long as the norm of the latent is around $20$. Thus, in this work, we plan to normalize every reconstructed latent $\hat{\bz}$ to this norm before generating any images with UnCLIP. In the following, we consider all the reconstructed latent vectors $\hat{\bz}$ to have already been normalized for generation.

    \begin{figure*}[htbp]
        \centering
        \includegraphics[width=\textwidth]{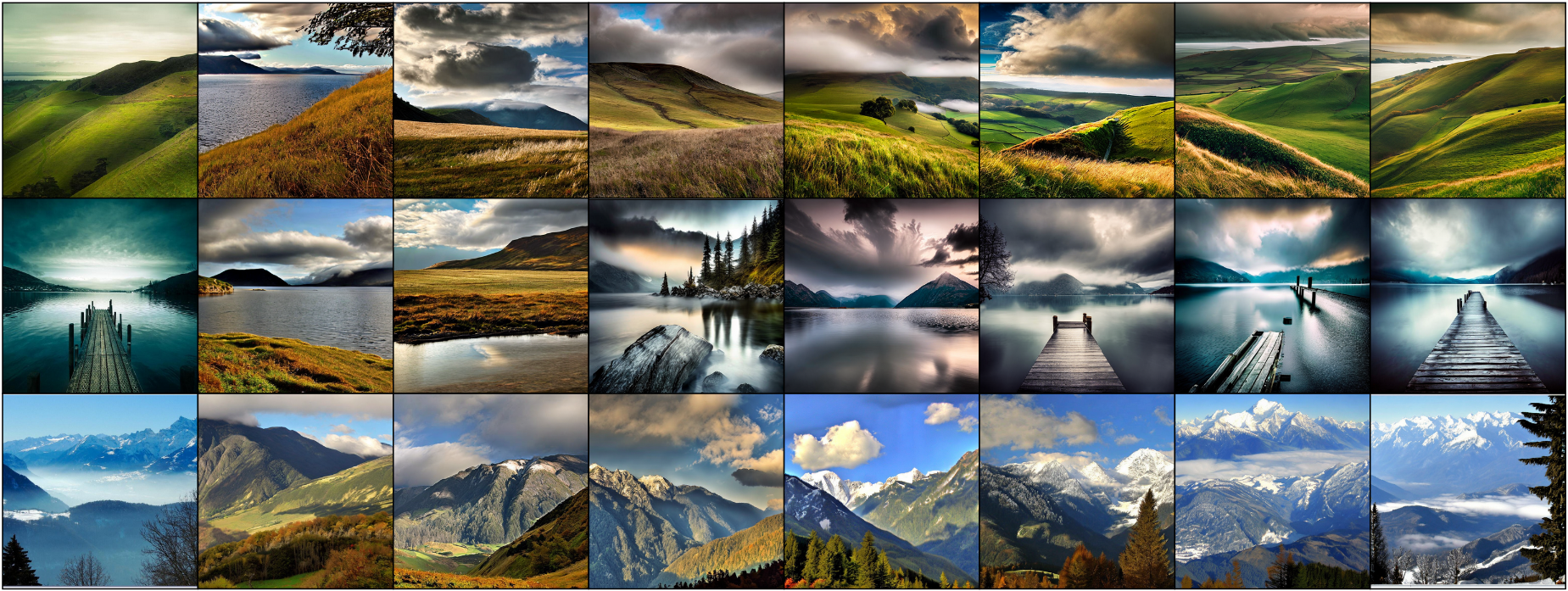}
        \caption{Generated images from their dictionary projection. The dictionary is learned on \cite{landscape}. (Left to right) Input images. Generated images for $n_a\in [2, 4, 8, 16, 32, 64, 128]$}
        \label{fig:recon_ex}
    \end{figure*}

    \cite{bachard2024} also shows that a CLIP-UnCLIP-based compression format conserves the semantics of the images. We also have to ensure that, in SMIC, the semantic of the images are conserved when generating images, especially through the dictionary reconstruction. Fig.~\ref{fig:recon_ex} shows images generated from different input images and with dictionaries of different sizes. For these dictionaries, $n_a\in[2,4, 8, 16, 32, 64, 128]$ and $\lambda=0.5$. We first observe that the generated images are more and more semantically close to their respective inputs as $n_a$ grows. Indeed, as the dictionary becomes larger, more specific semantics can be extracted from the image collection, and the projection becomes more precise. We can interpret this as the atoms of the dictionary being like semantic frequencies: when only a few of them are allowed (small $n_a$), only rough and general semantics is present, to grasp the maximum information about the database. When $n_a$ increases, more and more details are available in the dictionary to semantically reconstruct the inputs, like high frequencies. This can be observed with the general style of the generated images, the higher $n_a$, the closer the style; or more specifically with the second example, where the bridge on the lake is generated (and thus present in the dictionary) only when $n_a \leqslant 32$. The second observation from this experiment is that even at very low values of $n_a$, we can notice semantic differences in the generated images. For example, at only $n_a=4$, we can differentiate the plains from a lake (even though other details are present) from the mountains. This observation is confirmed in Table.~\ref{tab:sem_atom}, where we clearly notice a positive correlation between the number of atoms $n_a$ and the semantic metrics $d_{\Phi}$.

    \begin{table}[htpb]
        \centering
        \begin{tabular}{ccccc}
            \toprule
            $n_a$ & CC $\nearrow$ & CSS $\nearrow$ & BSS $\nearrow$ & MSE-SS $\searrow$  \\
            \midrule
            \cite{bachard2024}  & $0.891$ & $0.751$ & $0.512$ & $-$     \\
            $2$                 & $0.806$ & $0.525$ & $0.393$ & $0.545$ \\
            $4$                 & $0.822$ & $0.558$ & $0.408$ & $0.539$ \\
            $8$                 & $0.835$ & $0.619$ & $0.457$ & $0.488$ \\
            $16$                & $0.859$ & $0.680$ & $0.473$ & $0.431$ \\
            $32$                & $0.873$ & $0.708$ & $0.486$ & $0.408$ \\
            $64$                & $0.881$ & $0.712$ & $0.501$ & $0.385$ \\
            $128$               & $0.893$ & $0.732$ & $0.516$ & $0.382$ \\
            \bottomrule
        \end{tabular}
        \ \\
        \caption{Evolution of semantic coherence regarding $n_a$.}
        \label{tab:sem_atom}
    \end{table}

    All in all, we demonstrate that solving Eq.~(\ref{eq:learn_dict}) gives us a semantic dictionary $\dico$ over $\mathcal{X}$, and that this dictionary can be used to reconstruct the latent vectors of the images in the data collection and then generate qualitative images that are semantically coherent with their respective inputs.

    \begin{figure*}[hbtp]
        \centering

        \includegraphics[width=\textwidth]{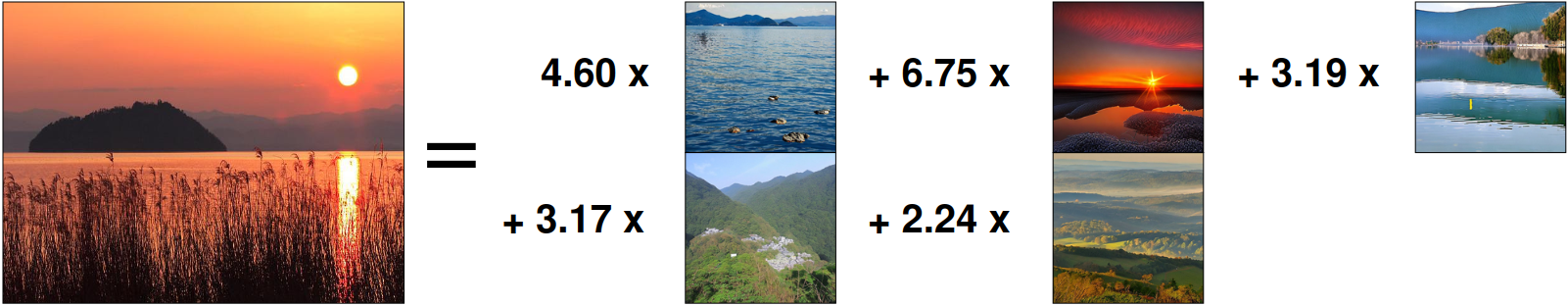}
        \caption{Example of decomposition in learned dictionary with $\alpha =0.75$ and $n_a=32$. (Left) Input image. (Right) Non-null atom and their associated coefficient. }
        \label{fig:decomp}
    \end{figure*}

    To get a better grasp of the semantic nature of the dictionary, we looked at the decomposition of images in the learned dictionary. An example is proposed in Fig.~\ref{fig:decomp}, and other examples are detailed in the supplementary materials. We observe through these examples that the atoms which coefficients are non-zero in the decomposition are semantically coherent with the semantic of the input. The input image of Fig.~\ref{fig:decomp} is clearly decomposed into the sum of its semantics components, with more emphasis on the most representative ones: \say{sea} + \say{sunset} + \say{lake} + \say{mountain} + \say{mountain}.  

    \subsection{Discussion}

    In this section, we push the experiments beyond the initial frame of semantic compression. These experiments will help to grasp a more profound understanding of the semantic properties of the dictionaries and their atoms.

    The previous experiment demonstrated the semantic expressiveness of the learned dictionaries over the images' collection. In this section, we explore some limitations of these semantic dictionary-based reconstruction methods. More specifically, we explore the semantic of generated images which initial inputs are semantically outside the data collection used to learn the dictionary.

\begin{figure*}
        \centering
        \includegraphics[width=\textwidth]{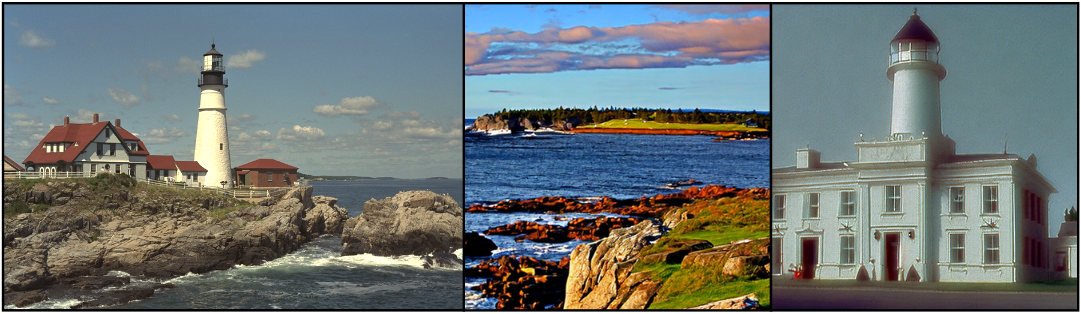}
        \caption{Image generated from the projection and the residual of an image from \cite{kodak} with a dictionary learned on \cite{landscape}. (Left) Input image. (Middle) Image generated via $\hat{\bz}$. (Right) Image generated via $\bar{\bz}$.}
        \label{fig:proj_resi}
    \end{figure*}

    Fig.~\ref{fig:proj_resi} shows an input image that is not part of the Landscape dataset, and what UnCLIP generates if we still try to solve Eq.~(\ref{eq:learn_coef}) for the reconstructing the latent vector. To complete this observation, we also generate images from the residual $\bar{\bz} = \bz - \hat{\bz}$ and we evaluate the semantics. Note that every latent vector, both $\hat{\bz}$ and $\bar{\bz}$, are normalized to $20$, as recommended by \cite{bachard2024}. More example are available in the supplementary materials.

    From Fig.~\ref{fig:proj_resi} we observe an interesting semantic separation. Indeed, the images generated from the projection express semantics that are correlated to the semantics of the database from which the dictionary has been learned. In this example, the images generated depict landscapes. On the other hand, the images generated with the residuals of the projection and the input latent only express semantic content that is absent from the database used to learn the dictionary. This property shows that both the projection and the residuals contain relevant semantic information and that a semantic dictionary learned of specific data can be used as a semantic filter, at least for image generation. This can lead to quantization algorithms that are specific to semantic quantization. However, these algorithms would be outside the scope of this work and are left as future work. Finally, note that the sum of all coefficients is around $20$, as expected by \cite{bachard2024}. Indeed, as all the atoms of dictionaries are exactly $1$, is it expected that their sum is around the mean of the norms.

\section{Semantic Multi-Item Compression}

    In this section, we propose a multi-item generative compression framework based on CLIP and on learning a dictionary, Eq.~(\ref{eq:learn_dict}), to grasp a description of the latent space spanned by the database. As there are multiple parameters to tune with the proposed framework, we then proceed to study their impacts through rate-distortion optimization. Finally, we compare our framework to state-of-the-art compression pipelines and show that the proposed compression scheme beats them both in terms of semantic conservation and in compression rates, even with the dictionary overhead. 
    
    \subsection{Coding scheme}

\begin{figure*}[htbp]
        \centering
        \includegraphics[width=\textwidth]{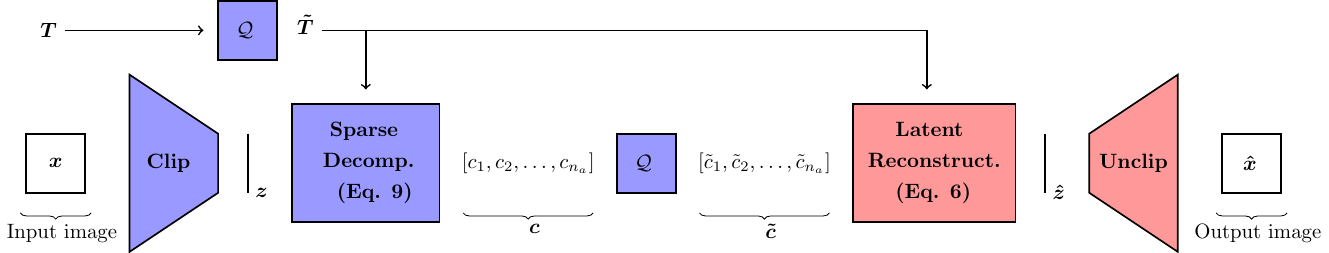}
        \caption{Individual MIGC compression pipeline. To compress a database, this needs to be done on every image.}
        \label{fig:enc-dec}
    \end{figure*}

    From the semantic properties of CLIP, \cite{bachard2024} and Sec. III, and from the capacity of learning a dictionary that can encapsulate the CLIP region spanned by the semantics of a database, we propose a multi-item generative compression framework. Fig.~\ref{fig:learn-dict} and \ref{fig:enc-dec} depict the two-step proposed framework to propose the generative multi-item compression of the database $\mathcal{X}$. 
    
    The first step, presented in Fig.~\ref{fig:learn-dict}, is to grasp the useful statistics of the database, $\dbi$. Following the pipeline of Sec. VI, we encode all the images of the database $\mathcal{X}$ into CLIP's latent space, represented with an aggregated matrix $\bm{Z}$. Then, we learn a semantic latent dictionary $\dico$ by solving Eq.~(\ref{eq:learn_dict}). Finally, as the dictionary needs to be transmitted and can then represent a bottleneck, we also quantize each atom alongside each dimension through uniform quantization to a fixed number of bits $b_{\text{dict}}$ into a quantized dictionary $\dicoq$. Indeed, both prior experiments and \cite{bachard2024} point toward, at worst, minimal impact on the compression scheme. The rate of these databases' statistics is $\mathcal{R}(\dicoq)$.

    The second step, the individual coding of an image of the database, is presented in Fig.~\ref{fig:enc-dec}. First, the image $\bx$ is encoded into a CLIP latent vector $\bz$. This latent vector is then transformed into coefficients $\bm{c}$ by solving Eq.~(\ref{eq:learn_coef}) with the previously learned $\dicoq$. As we propose a dictionary-based compression scheme, we expect $\bm{c}$ to be sparse, and its rate is approximated by the number of non-zero coefficients. To achieve better compression gains, each coefficient is then quantized using uniform quantization to a fixed number of bits $b_{\text{coef}}$ into $\tilde{\bm{c}}$, for a rate $\mathcal{R}(\tilde{\bm{c}})$. 

    To decode an image, we solve Eq.~(\ref{eq:decomposition}) to reconstruct the initial latent vector by using the transmitted semantic dictionary $\dicoq$ into $\hat{\bz}$. Finally, UnCLIP is used to generate images $\hat{\bx}$ from $\hat{\bz}$ that have the same semantics as the original input image $\bx$. Note that with the proposed compression scheme, we can encode and decode an image independently of the rest of the database. The random access property, typical of single-item compression, is preserved.

    The impact of each parameter, $b_{\text{dict}}$ and $b_{\text{coef}}$, but also the sparsity of the decomposition or the number of atoms, is discussed in the next section. All in all, the problem solved, given a semantic threshold $\tau_{\Phi}$, is proposed in the following equation:

    \begin{gather} \label{eq:final_pb}
        \min \sum_{i=1}^N \mathcal{R}(\tilde{\bm{c}_i}) + \mathcal{R}(\dicoq) \ \text{  s. t. } \\
        \forall i\in \llbracket 1, N \rrbracket,\ \Phi\left(\bx_i,\ \hat{\bx}_i\right ) < \tau_{\Phi} \nonumber
    \end{gather}

    Summary of the compression scheme:
    \begin{itemize}
        \item Learn a semantic dictionary $\dico$ from a collection of images $\mathcal{X}$ by solving Eq.~(\ref{eq:learn_dict});
        \item Quantize $\dicoq$;
        \item Encode an image $\bx$ by projecting its CLIP latent vector $\bz$ onto a list of coefficients $\bm{c}$ using $\dicoq$;
        \item Quantize $\bm{c}$;
        \item Decode an image by reconstructing the CLIP latent vector using $\dicoq$ and then generate images $\hat{\bx}$ via UnCLIP.
    \end{itemize}

    \subsection{Compression rate of a data collection}
    \label{sec:comprate}

    The advantage of multi-item compression is that it considers statistics and redundancies in a database to perform a better item-wise compression. However, considering this information comes with an additional cost $\dbi$, in this case $\mathcal{R}(\dicoq)$, that has to be accounted for. In this section, we calculate the rate of the whole compressed database and the dictionary over cost, given the different hyperparameters of the pipeline.

    From Eq.~(\ref{eq:final_pb}), we have that the total rate of compression for a database $\mathcal{R}_{total}$ is given by:

    \begin{align}
        \mathcal{R}_{total} = \mathcal{R}(\dbi) + N *\mathcal{R}(\bz)
    \end{align}
    Where $\mathcal{R}(\bz)$ is the expected compression rate of a latent vector over the whole database.
    
    We can now explicitly define the right-hand terms in terms of the parameters discussed in the previous sections. The database side information, $\mathcal{R}(\dbi)$ is given by:
    \begin{align}
        \mathcal{R}(\dbi) = \mathcal{R}(\dicoq) = n_a * \LS *\ b_{\text{dict}}
    \end{align}
    And for $|\bz|$, with the proposed coding scheme for the coefficients in the previous section:
    \begin{align}
        \mathcal{R}(\bz) = \log_2(n_a) * b_c * P(c\text{ is non-null})
    \end{align}

    \begin{figure}[htbp]
        \centering
        \includegraphics[width=.5\textwidth]{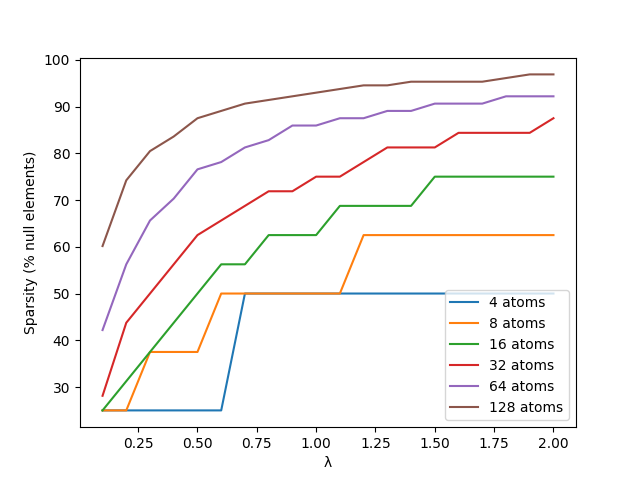}
        \caption{Proportion of null elements in the coefficients' list.}
        \label{fig:p_null}
    \end{figure}
    
    Furthermore, a study of the evolution of $P(\bm{c}\text{ is null})$ as a function of $\lambda$ and $n_a$ is proposed in Fig.~\ref{fig:p_null}. From this figure, we derive the following equation for the proportion of non-null coefficients in the coefficients' list:
    \begin{align}
        P(\bm{c}\text{ is null}) = 1 - P(\bm{c}\text{ is non-null}) \approx \frac{1}{(\lambda + 1)^{\log_2(n_a)}}
    \end{align}

    Finally, the total compression rate of the compressed database, taking into account the dictionary overcost, is:
    \begin{align}
        \mathcal{R}_{total} = n_a * \LS *\ b_{\text{dict}} + \frac{N * \log_2(n_a) * b_c}{(\lambda + 1)^{\log_2(n_a)}} 
    \end{align}

    \subsection{Rate Semantic Fidelity Optimization}

    Given the problem we are solving, Eq.~(\ref{eq:final_pb}), and the proposed compression pipeline, we are looking for the best values for the different parameters involved. To do so, we proposed to first grasp the impact of each parameter on its own, with the others fixed to decent values. Next, we evaluate every possible set of parameters through the compression rate and how much the semantic is conserved. Finally, we compute the upper part of the convex hull of these results, as these parameters represent the best parameters for a given rate-semantic distortion trade-off. 

    In this work, we evaluate and discuss the impact of the following parameters:
    \begin{itemize}
        \item $n_a$, the number of atoms in the dictionary;
        \item $\lambda$, the sparsity of the coefficient in Eq.~\ref{eq:learn_coef};
        \item $b_{\text{dict}}$, the number of bits per dimension for each atom of the dictionary; 
        \item $b_{\text{coef}}$, the number of bits per coefficient.
    \end{itemize}
    For this study, to study a specific parameter, if needed, we arbitrarily fix the other parameters to $n_a=64$, $\lambda=1$, $b_{\text{dict}}=16$, $b_{\text{coeff}}=16$.

    \textbf{Impact of the sparsity of the coefficients:} The average non-null number of coefficients in a dictionary-based compression pipeline is a strong bottleneck, as it can change the way of how to encode the coefficients. The study linking $\lambda$ and the proportion of non-null coefficients is proposed in Fig.~\ref{fig:p_null}. From this experiment, we observe that the proportion (or probability) that a coefficient is null is increasing with $\lambda$, with a rough $\frac{1}{(1+\lambda)^{\log{_2}(n_a)}}$ fashion (where the exponent of $(1+\lambda)$ is debatable). Now that we have a clear link between $\lambda$ and the proportion of non-null coefficients, we can link the semantic conservation to $\lambda$ to highlight the expected trade-off between semantic conservation and rate. We observe a clear trade-off, as $\lambda$ increases, all the semantic-based metrics decrease (figures presented in the supplementary materials). 

    This discussion made us choose to encode the coefficients as a set of tuples, where one is the (possibly) quantized coefficient, and the other is the number of the associated atoms. This way, we opt for a medium-to-high sparse representation of the coefficients, as the additional cost of the position will be negligible in front of the number of zeros not encoded. 

    \textbf{Impact of the rest of the parameters on the conservation of the semantic:} The evolution of the semantic coherence evolves as expected, the higher the value of a parameter, the better the conservation of the semantic. However, each parameter does not have the same impact on the semantic fidelity. From the experiments and the figures given in the supplementary material, we deduce the following importance ranking: $\lambda >>> n_a > d_{\text{coef}} >>> d_{\text{dico}}$. The first parameter to change is $\lambda$; as both the bitrate and the semantic metric increase, $\lambda$ decreases. Indeed, as the sparsity of the coefficients decreases, more information is transmitted, which allows for better semantic coherence in the pipeline. Then, when $\lambda$ reaches $0$, other parameters start to evolve, and $\lambda$ resets to $2$. We first observe an increase in the size of the dictionary $n_a$, followed by an increase in the bit rate of the coefficients $b_{\text{coef}}$. Finally, the last parameter to increase is the bit rate of the atoms, $b_{\text{dico}}$. Indeed, as expected by the study of the parameter and by the results in \cite{bachard2024}, CLIP latent vectors can be quantized in a very harsh way and still keep their semantic content. We can argue that the semantic addition of increasing $b_{\text{dico}}$ is marginal but non-negligible, at the expense of a considerable increase in the bits that have to be transmitted.

\begin{figure*}[htbp]
        \centering
        \includegraphics[width=0.45\textwidth]{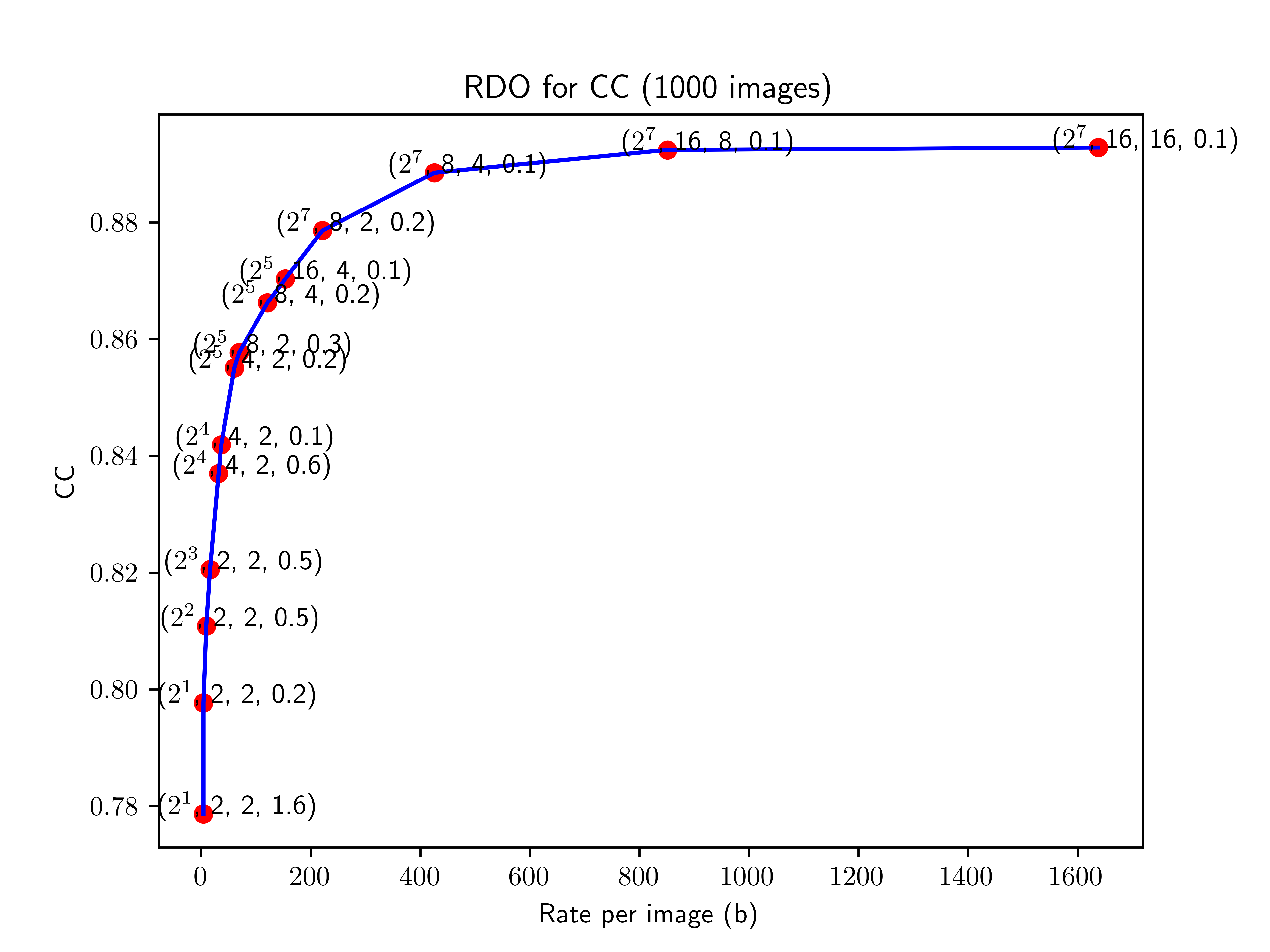}
        \hfill
        \includegraphics[width=0.45\textwidth]{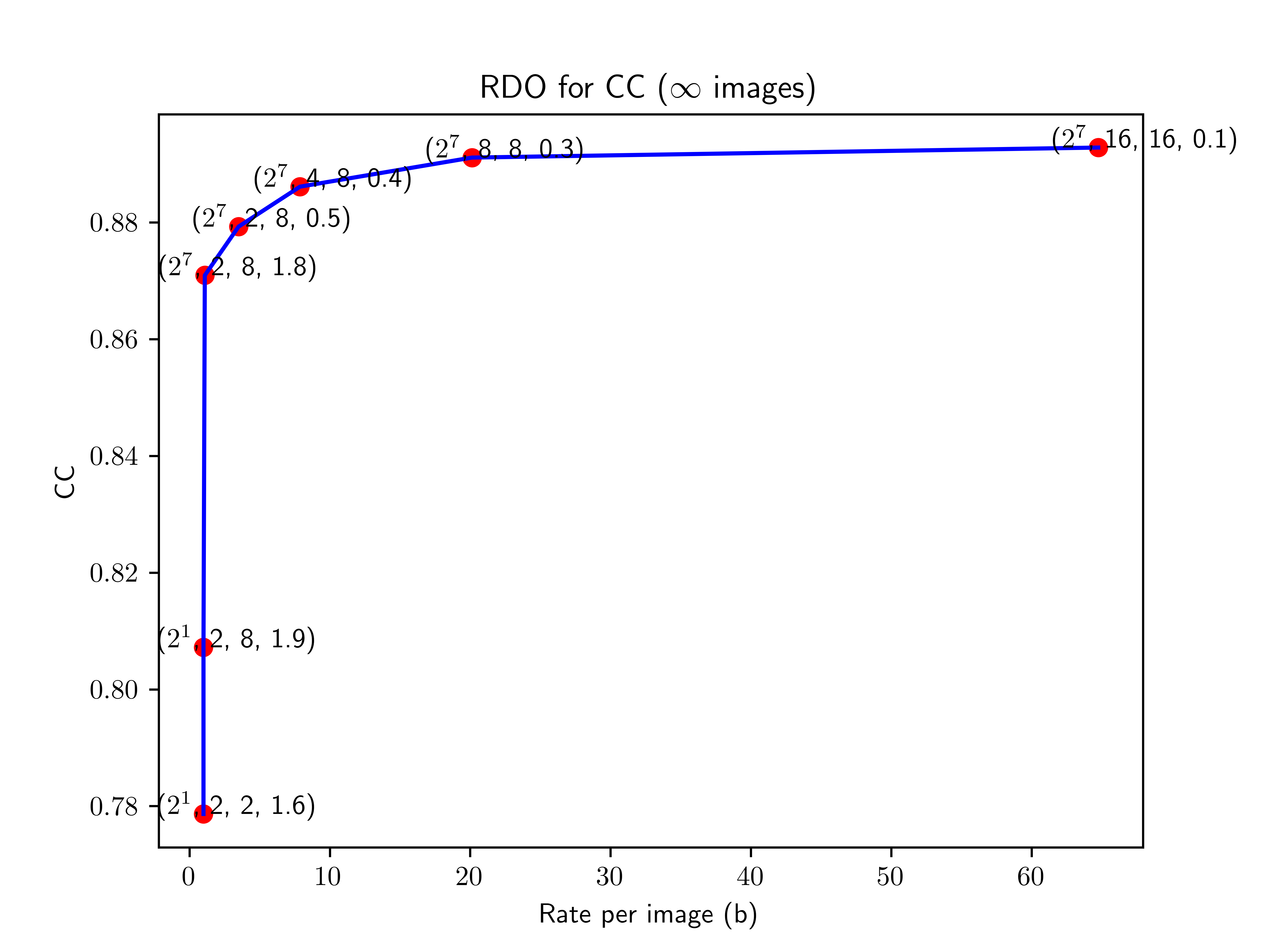}
        \caption{Rate-distortion optimization curves for CC and the associated $(n_a, b_{\text{coef}}, b_{\text{dico}}, \lambda)$ parameter sets. (Left) $n=1000$. (Right) $n=\infty$. More figures in the supplementary material.}
        \label{fig:rdo}
    \end{figure*}

    \textbf{Rate-semantic fidelity validation:} We showed that increasing the values of the parameters (or decreasing $\lambda$) increases the semantic conversation in the pipeline. However, the discussions in Sec.~\ref{sec:comprate} show that they also have an impact on the rate of the compressed database, as the rate increases with the increase in the parameters (diminution for $\lambda$), as expected by \cite{sc_rd}. To select the best sets of parameters, we conduct rate-semantic fidelity optimization (R-SF-O). Because we deal with multi-item compression, we have to take into account the overhead of the data collection's statistics (here the semantic dictionary), so a classical R-SF-O is not possible. To overcome this difficulty, we propose an R-SF-O for different sizes of the database $n$, and the rate is expressed in bits per image (here all the images are $768\times768$) while considering the dictionary overhead. R-SF-O are proposed in Fig.~\ref{fig:rdo} for $n=100, 10000, 100000$ and $n=\infty$ (where the overhead of the dictionary is completely absorbed). On these figures, we also plotted in blue the upper part of the convex hull of the different experiments. These points represent the best set of parameters for a given rate-distortion trade-off. 

    From Fig.~\ref{fig:rdo} we observe, for the proposed framework, the classical rate-distortion trade-off, even for semantics metrics: you either minimize the rate or maximize the metrics. This leads us to several sets of parameters, depending on $n$, that can be used for practical image collection compression. From these tables, we observe that the size of the database $n$ has an impact on the best sets of parameters. Indeed, as the metric scores do not change, the rate moves non-linearly, thus spreading and ordering the different experiments points differently depending on $n$. Second, we also observe that the full range of parameters ($n_a, b_{\text{dico}}, b_{\text{coef}}, \lambda$) are used as best parameters for a given rate-distortion trade-off. 

    For future comparisons, we select the parameters from the results of the experiment where $n=10000$, as there are around $5000$ images in the Landscape \cite{landscape} database.

    \subsection{Comparison to the state-of-the-art}

    In this section, we compare the proposed compression scheme to state-of-the-art compression algorithms. Because inter-item compression frameworks only consider pixel-redundancies, and not higher-level redundancies, we mainly focus on comparing with single-item compression schemes. First, we compare our work with its single-item version \cite{bachard2022} to select the best parameter set. Regarding the discussion in the previous section, we define $3$ models from this framework:
    \begin{description}
        \item[\emph{low}:] \hspace{.3cm} $n_a=2^1, b_{\text{coef}}=2^1, b_{\text{dico}}=2^1, \lambda=1.6$
        \item[\emph{medium}:] \hspace{.3cm} $n_a=2^7, b_{\text{coef}}=2^2, b_{\text{dico}}=2^2, \lambda=0.2$
        \item[\emph{high}:] \hspace{.3cm} $n_a=2^7, b_{\text{coef}}=2^4, b_{\text{dico}}=2^4, \lambda=0.1$
    \end{description}
    An example of images decoded with each of these codecs is proposed in Fig.~\ref{fig:ex_models} and more are available in the supplementary materials.

    \begin{figure*}
        \centering
        \includegraphics[width=\linewidth]{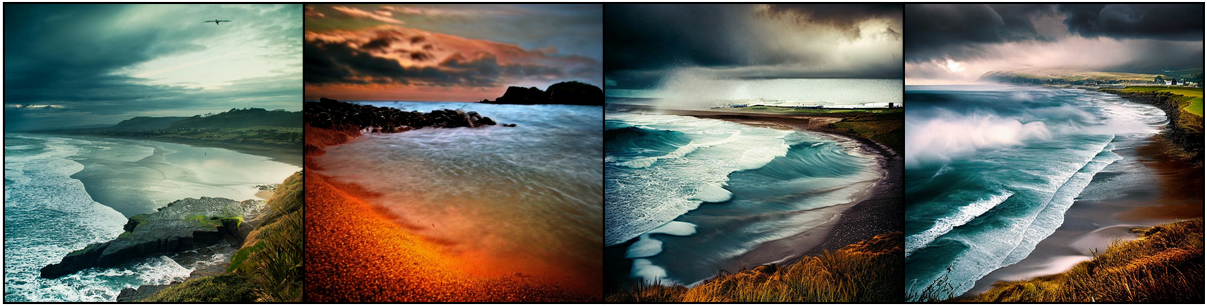}
        \caption{Images generated from our different models. (Left to right) Input image. Image generated via the \emph{low} model, via the \emph{medium} model, and via the \emph{high} model}
        \label{fig:ex_models}
    \end{figure*}

    To compare our models to single-item compression scheme, we define $n^*$, the minimal size $n^*$ of a database for which the proposed SMIC scheme is more interesting than encoding all the items with a SIC scheme. The formula is the following one:
    \begin{gather}
        \mathcal{R}(\dbi) + n^{*} * \mathcal{R}(\bz_{SMIC}) < n^{*} * \mathcal{R}(\bz_{SIC}) \\ \nonumber
        n^* > \frac{\mathcal{R}(\dicoq)}{\mathcal{R}(\bz_{SIC}) - \mathcal{R}(\tilde{\bm{c}})}
    \end{gather}
    
    Table~\ref{tab:res_intra_sem} presents the (fixed) semantic conservation of the different models, and Table~\ref{tab:res_intra_bpp} presents the ratio $\frac{n
    \mathcal{R}(\bz_{SIC})}{\mathcal{R}(\dicoq) + n\mathcal{R}(\tilde{\bm{c}})}$ for different scenarios ($n$ and the models varying), as well as the computation for $n^*$ between ours models and their single item counterpart \cite{bachard2024}. From these results, we observe a trade-off between semantic conservation and compression rates. The images generated via the \emph{low} model lose some important semantic details (see the supplementary material). On the other hand, we also observe that the \emph{high} model is better than its single item counterpart only for huge data collections. We conclude that the medium model is a good trade-off between semantics preservation and compression gains, and we set this set of parameters as our model in the following.

    \begin{table}[htbp]
        \centering
        \begin{tabular}{cccc}
             \toprule
             Clip version & CC & BSS & CSS \\
             \midrule
             Inter (low) & $0.78$ & $0.38$ & $0.51$ \\ 
             Inter (medium) & $0.85$ & $0.49$ & $0.67$ \\
             Inter (high) & $0.89$ & $0.52$ & $0.74$ \\
             Intra (\cite{bachard2024}) & $0.90$ & $0.48$ & $0.72$ \\
             \bottomrule
        \end{tabular}
        \caption{Semantic conservation comparisons for Clip-based compression schemes. Different parametrizations of our inter-item model are compared to their intra version \cite{bachard2024}.}
        \label{tab:res_intra_sem}
    \end{table}

    \begin{table}[htbp]
        \centering
        \begin{tabular}{cccc}
             \toprule
             Size of the collection & Inter (low) & Inter (medium) & Inter (high) \\
             \midrule
             $n=100$ & $\boldsymbol{22.51}$ & $0.17$ &$0.04$ \\ 
             $n=1000$ & $\boldsymbol{184.3}$ & $\boldsymbol{1.76}$ & $0.43$ \\
             $n=10000$ & $\boldsymbol{659}$ & $\boldsymbol{15}$ & $\boldsymbol{3.3}$ \\
             $n=\infty$ & $\boldsymbol{923}$ & $\boldsymbol{90.9}$ & $\boldsymbol{12.3}$ \\
             $n^*$ with \cite{bachard2024} & $5$ & $562$ & $2419$ \\ 
             \bottomrule
        \end{tabular}
        \caption{Compression ratios between SIC (where $\mathcal{R}_{\bz_{SIC}}=1.2\times10^{-3}$ \cite{bachard2024}) and our SMIC models. $n=\infty$ means that the dictionary overhead is not considered. Bold results, greater than $1$, show when our model outperforms its single item counterpart.}
        \label{tab:res_intra_bpp}
    \end{table}

    \begin{table}[htbp]
        \centering
        \begin{tabular}{ccccccc}
            \toprule
            Model                           & Rate per image (BPP) & $n^*$ Compared to SMIC \\
            \midrule
            Clip inter (ours)               & $1.4\times10^{-4}$ & -- \\
            Clip intra (\cite{bachard2024}) & $1.2\times10^{-3}$ & $150$ \\
            Pics \cite{pics}                & $2.6\times10^{-2}$ & $5$ \\
            VVC \cite{VVC}                  & $4.5\times10^{-3}$ & $38$ \\
            \bottomrule
        \end{tabular}
        \caption{Bitrates comparison of different coding schemes. Ours is SMIC, others are SIG. The rate per image takes into account the overhead of the dictionary for our model. We set $n=5000$ in this comparison as it is the size of the Landscape data collection.}
        \label{tab:res}
    \end{table}

    Table \ref{tab:res} presents the comparisons made between our selected model, the \emph{medium} one, and \cite{bachard2024}, \cite{pics} and \cite{VVC} in terms of compression rates, and when the minimal database's size for when SMIC is more interesting than SIC if we encode the whole Landscape data collection. 
    
    From the results, we observe that our model beats the state-of-the-art SIC schemes in terms of compression costs with good semantic conservation (see \cite{bachard2024} for comparisons). Even with the over-cost of the dictionary, the database size $n^*$ from which MIGC is more interesting than SIC is $150$ for the most compact SIC framework (Clip intra) and less than $50$ otherwise. This indicates that the proposed scheme does not need enormous databases to be more efficient than SIC-based schemes; in some cases only a few images suffice. 

\section{Conclusion and Future Work}

    In this work, we demonstrated that the latent space induced by CLIP has semantic linearity properties. In short, one can add or subtract high-level concepts with classical additions or subtractions in the latent space seen as a $\mathbb{R}$ vector space. From these properties, we derived a multi-item dictionary-based compression scheme that beats state-of-the-art in terms of compression, even with the over-cost of the dictionary, for databases that are made of a few hundred images or more.

    Moreover, we showed that the learned dictionaries can be used as a projection basis for separating the semantic of images.

    From the examples of separating the semantics of images into the semantics of the database and the semantics outside can lead to the definition of a family of small dictionaries. Each dictionary would describe a very precise semantic relative to the task. From these sets of dictionaries, one could then derive a semantic-based quantization algorithm based on the importance of some semantics concepts over others. This kind of semantic-based quantization may be more adapted to user-based quantization than the classical uniform quantization proposed in this work.

    Another possible adaptation of this work can be to look for other foundation models that have easy semantic separation properties and see how to exploit them for compression or even for other semantic related tasks. 

    \bibliographystyle{IEEEtran}
    \bibliography{bib}

% Generated by IEEEtran.bst, version: 1.14 (2015/08/26)
\begin{thebibliography}{10}
\providecommand{\url}[1]{#1}
\csname url@samestyle\endcsname
\providecommand{\newblock}{\relax}
\providecommand{\bibinfo}[2]{#2}
\providecommand{\BIBentrySTDinterwordspacing}{\spaceskip=0pt\relax}
\providecommand{\BIBentryALTinterwordstretchfactor}{4}
\providecommand{\BIBentryALTinterwordspacing}{\spaceskip=\fontdimen2\font plus
\BIBentryALTinterwordstretchfactor\fontdimen3\font minus \fontdimen4\font\relax}
\providecommand{\BIBforeignlanguage}[2]{{%
\expandafter\ifx\csname l@#1\endcsname\relax
\typeout{** WARNING: IEEEtran.bst: No hyphenation pattern has been}%
\typeout{** loaded for the language `#1'. Using the pattern for}%
\typeout{** the default language instead.}%
\else
\language=\csname l@#1\endcsname
\fi
#2}}
\providecommand{\BIBdecl}{\relax}
\BIBdecl

\bibitem{bross2021overview}
B.~Bross, Y.-K. Wang, Y.~Ye, S.~Liu, J.~Chen, G.~J. Sullivan, and J.-R. Ohm, ``Overview of the versatile video coding (vvc) standard and its applications,'' \emph{IEEE Transactions on Circuits and Systems for Video Technology}, vol.~31, no.~10, pp. 3736--3764, 2021.

\bibitem{girod1997performance}
B.~Girod, E.~Steinbach, and N.~F{\"a}rber, ``Performance of the h. 263 video compression standard,'' \emph{Journal of VLSI signal processing systems for signal, image and video technology}, vol.~17, pp. 101--111, 1997.

\bibitem{richardson2011h}
I.~E. Richardson, \emph{The H. 264 advanced video compression standard}.\hskip 1em plus 0.5em minus 0.4em\relax John Wiley \& Sons, 2011.

\bibitem{vanne2012comparative}
J.~Vanne, M.~Viitanen, T.~D. Hamalainen, and A.~Hallapuro, ``Comparative rate-distortion-complexity analysis of hevc and avc video codecs,'' \emph{IEEE Transactions on Circuits and Systems for Video Technology}, vol.~22, no.~12, pp. 1885--1898, 2012.

\bibitem{webdomo}
\BIBentryALTinterwordspacing
``https://www.domo.com/learn/infographic/data-never-sleeps-11.'' [Online]. Available: \url{https://www.domo.com/learn/infographic/data-never-sleeps-9}
\BIBentrySTDinterwordspacing

\bibitem{MIC1}
H.~Wu, X.~Sun, J.~Yang, W.~Zeng, and F.~Wu, ``Lossless compression of jpeg coded photo collections,'' \emph{IEEE Transactions on Image Processing}, vol.~25, no.~6, pp. 2684--2696, 2016.

\bibitem{MIC2}
P.~M. Latha and A.~A. Fathima, ``Collective compression of images using averaging and transform coding,'' \emph{Measurement}, vol. 135, pp. 795--805, 2019.

\bibitem{MIC3}
L.~Sha, W.~Wu, and B.~Li, ``Novel image set compression algorithm using rate-distortion optimized multiple reference image selection,'' \emph{IEEE Access}, vol.~6, pp. 66\,903--66\,913, 2018.

\bibitem{MIC4}
\BIBentryALTinterwordspacing
J.~Luo, S.~Li, W.~Dai, C.~Li, J.~Zou, and H.~Xiong, ``Learned lossless compression for jpeg via frequency-domain prediction,'' 2023. [Online]. Available: \url{https://arxiv.org/abs/2303.02666}
\BIBentrySTDinterwordspacing

\bibitem{MIC5}
L.~Sha, W.~Wu, and B.~Li, ``Low-complexity and high-coding-efficiency image deletion for compressed image sets in cloud servers,'' \emph{IEEE Transactions on Cloud Computing}, vol.~11, no.~1, pp. 608--619, 2021.

\bibitem{MIC6}
\BIBentryALTinterwordspacing
------, ``Image set compression for similar images with priorities,'' \emph{Electronics Letters}, vol.~55, no.~5, pp. 262--264, 2019. [Online]. Available: \url{https://ietresearch.onlinelibrary.wiley.com/doi/abs/10.1049/el.2018.7342}
\BIBentrySTDinterwordspacing

\bibitem{MIC7}
H.~Wu, X.~Sun, J.~Yang, W.~Zeng, and F.~Wu, ``Lossless compression of jpeg coded photo collections,'' \emph{IEEE Transactions on Image Processing}, vol.~25, no.~6, pp. 2684--2696, 2016.

\bibitem{bachard2022}
T.~Bachard, A.~J. Tom, and T.~Maugey, ``Semantic alignment for multi-item compression,'' in \emph{2022 IEEE International Conference on Image Processing (ICIP)}, 2022, pp. 2841--2845.

\bibitem{blau2017}
\BIBentryALTinterwordspacing
Y.~Blau and T.~Michaeli, ``The perception-distortion tradeoff,'' \emph{CoRR}, vol. abs/1711.06077, 2017. [Online]. Available: \url{http://arxiv.org/abs/1711.06077}
\BIBentrySTDinterwordspacing

\bibitem{chaudhuri2018decision}
A.~Chaudhuri, I.~Dukovska-Popovska, N.~Subramanian, H.~K. Chan, and R.~Bai, ``Decision-making in cold chain logistics using data analytics: a literature review,'' \emph{The International Journal of Logistics Management}, vol.~29, no.~3, pp. 839--861, 2018.

\bibitem{codingformachine}
N.~Le, H.~Zhang, F.~Cricri, R.~Ghaznavi-Youvalari, and E.~Rahtu, ``Image coding for machines: an end-to-end learned approach,'' in \emph{ICASSP 2021 - 2021 IEEE International Conference on Acoustics, Speech and Signal Processing (ICASSP)}, 2021, pp. 1590--1594.

\bibitem{llm}
\BIBentryALTinterwordspacing
H.~Gilbert, M.~Sandborn, D.~C. Schmidt, J.~Spencer-Smith, and J.~White, ``Semantic compression with large language models,'' 2023. [Online]. Available: \url{https://arxiv.org/abs/2304.12512}
\BIBentrySTDinterwordspacing

\bibitem{sc_net}
Z.~Hong, S.~Chen, G.-S. Xie, W.~Yang, J.~Zhao, Y.~Shao, Q.~Peng, and X.~You, ``Semantic compression embedding for generative zero-shot learning.'' in \emph{IJCAI}, 2022, pp. 956--963.

\bibitem{goodfellow2020generative}
I.~Goodfellow, J.~Pouget-Abadie, M.~Mirza, B.~Xu, D.~Warde-Farley, S.~Ozair, A.~Courville, and Y.~Bengio, ``Generative adversarial networks,'' \emph{Communications of the ACM}, vol.~63, no.~11, pp. 139--144, 2020.

\bibitem{ho2020}
\BIBentryALTinterwordspacing
J.~Ho, A.~Jain, and P.~Abbeel, ``Denoising diffusion probabilistic models,'' 2020. [Online]. Available: \url{https://arxiv.org/abs/2006.11239}
\BIBentrySTDinterwordspacing

\bibitem{bachard2024}
T.~Bachard and T.~Maugey, ``Can image compression rely on clip?'' \emph{IEEE Access}, 2024.

\bibitem{clip}
\BIBentryALTinterwordspacing
A.~Radford, J.~W. Kim, C.~Hallacy, A.~Ramesh, G.~Goh, S.~Agarwal, G.~Sastry, A.~Askell, P.~Mishkin, J.~Clark, G.~Krueger, and I.~Sutskever, ``Learning transferable visual models from natural language supervision,'' \emph{CoRR}, vol. abs/2103.00020, 2021. [Online]. Available: \url{https://arxiv.org/abs/2103.00020}
\BIBentrySTDinterwordspacing

\bibitem{unclip}
A.~Ramesh, P.~Dhariwal, A.~Nichol, C.~Chu, and M.~Chen, ``Hierarchical text-conditional image generation with clip latents,'' 2022.

\bibitem{stablediffusion}
R.~Rombach, A.~Blattmann, D.~Lorenz, P.~Esser, and B.~Ommer, ``High-resolution image synthesis with latent diffusion models,'' in \emph{Proceedings of the IEEE/CVF Conference on Computer Vision and Pattern Recognition (CVPR)}, 6 2022, pp. 10\,684--10\,695.

\bibitem{kodak}
\BIBentryALTinterwordspacing
Kodakt, ``Kodak lossless true color image suite,'' 1999. [Online]. Available: \url{https://r0k.us/graphics/kodak/}
\BIBentrySTDinterwordspacing

\bibitem{scikit}
J.~Mairal, F.~Bach, J.~Ponce, and G.~Sapiro, ``Online dictionary learning for sparse coding,'' in \emph{Proceedings of the 26th annual international conference on machine learning}, 2009, pp. 689--696.

\bibitem{landscape}
M.~Afifi, M.~A. Brubaker, and M.~S. Brown, ``Histogan: Controlling colors of gan-generated and real images via color histograms,'' in \emph{Proceedings of the IEEE Conference on Computer Vision and Pattern Recognition}, 2021.

\bibitem{clipscore}
J.~Hessel, A.~Holtzman, M.~Forbes, R.~L. Bras, and Y.~Choi, ``Clipscore: A reference-free evaluation metric for image captioning,'' \emph{arXiv preprint arXiv:2104.08718}, 2021.

\bibitem{lasso_cd}
\BIBentryALTinterwordspacing
T.~T. Wu and K.~Lange, ``Coordinate descent algorithms for lasso penalized regression,'' \emph{The Annals of Applied Statistics}, vol.~2, no.~1, 2008. [Online]. Available: \url{http://dx.doi.org/10.1214/07-AOAS147}
\BIBentrySTDinterwordspacing

\bibitem{sc_rd}
\BIBentryALTinterwordspacing
T.~Guo, Y.~Wang, J.~Han, H.~Wu, B.~Bai, and W.~Han, ``Semantic compression with side information: A rate-distortion perspective,'' 2022. [Online]. Available: \url{https://arxiv.org/abs/2208.06094}
\BIBentrySTDinterwordspacing

\bibitem{pics}
E.~Lei, Y.~B. Uslu, H.~Hassani, and S.~S. Bidokhti, ``Text+ sketch: Image compression at ultra low rates,'' \emph{arXiv preprint arXiv:2307.01944}, 2023.

\bibitem{VVC}
B.~Bross, J.~Chen, J.-R. Ohm, G.~J. Sullivan, and Y.-K. Wang, ``Developments in international video coding standardization after avc, with an overview of versatile video coding (vvc),'' \emph{Proceedings of the IEEE}, vol. 109, no.~9, pp. 1463--1493, 2021.

\end{thebibliography}

    \begin{IEEEbiography}[{\includegraphics[width=1in,height=1.25in,clip,keepaspectratio]{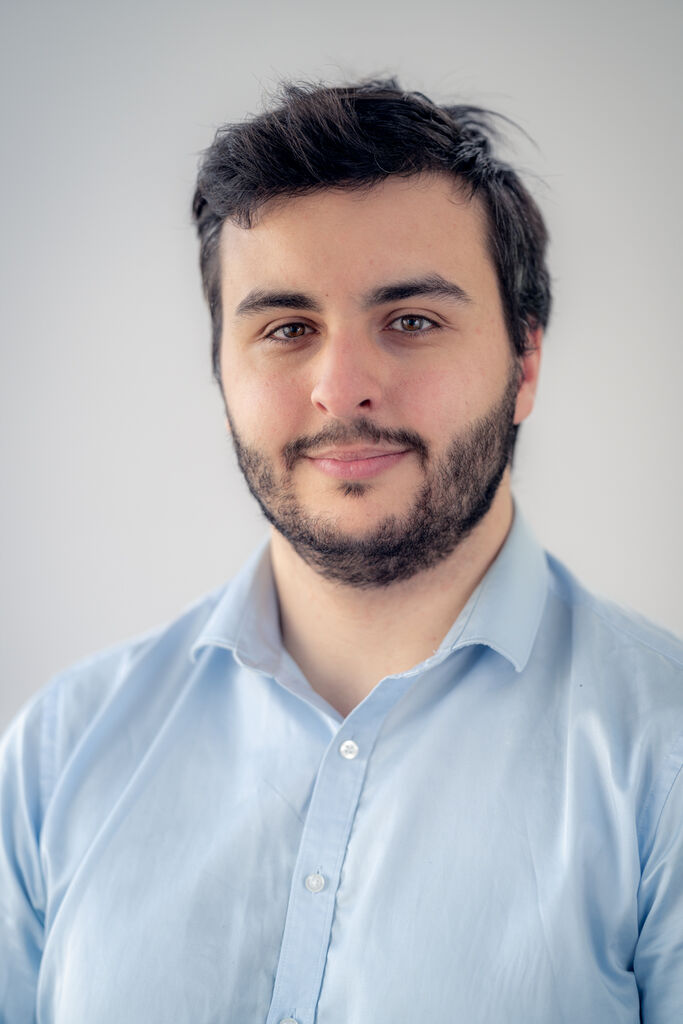}}]{Tom Bachard} Tom Bachard graduated from École Normale Supérieure de Rennes, Rennes, France, in 2021. He received an M.Sc. degree in Theoretical Computer Science from École Normale Supérieure de Rennes and Université Rennes 1, Rennes, France, in 2021. He started his PhD under the supervision of Thomas Maugey in 2021, at INRIA Bretagne, Rennes, France. His research interests lie in signal processing, image compression, and deep learning.
\end{IEEEbiography}

    \begin{IEEEbiography}[{\includegraphics[width=1in,height=1.25in,clip,keepaspectratio]{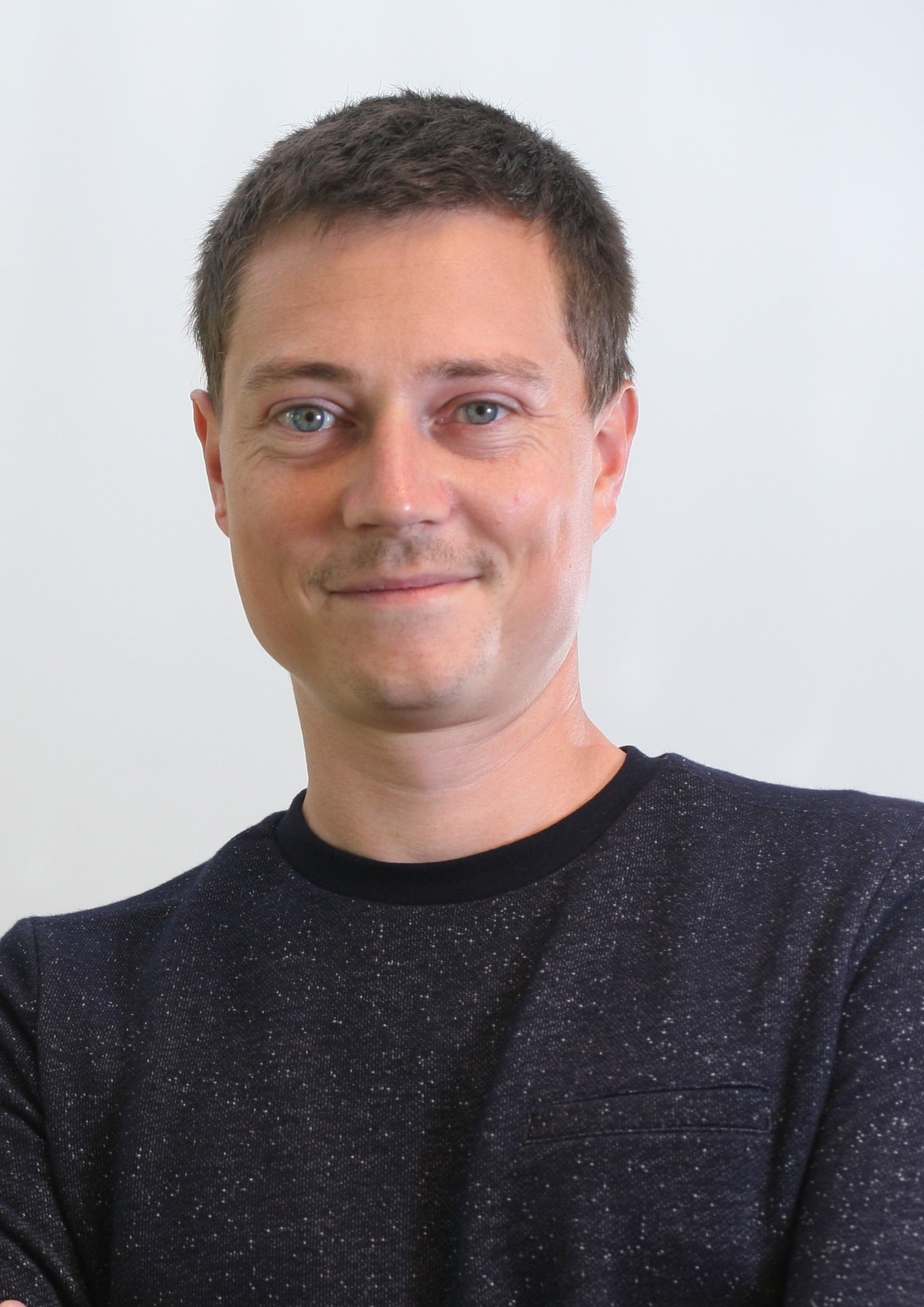}}]{Thomas Maugey} Thomas Maugey graduated from École Supérieure d’Electricité, Supélec, Gif-sur-Yvette, France in 2007. He received the M.Sc. degree in fundamental and applied mathematics from Supélec and Université Paul Verlaine, Metz, France, in 2007. He received his Ph.D. degree in Image and Signal Processing at TELECOM ParisTech, Paris, France, in 2010. From October 2010 to October 2014, he was a postdoctoral researcher at the Signal Processing Laboratory (LTS4) of the Swiss Federal Institute of Technology (EPFL), Lausanne, Switzerland. From November 2014 to October 2023, he was Research Scientist at Inria. Since October 2023, he has been Research Director at Inria. He serves as an Associate Editor for EURASIP Journal on  advances in signal processing and IEEE Signal Processing Letters. His research deals with image and video processing/compression and graph-based signal processing.
\end{IEEEbiography}

\end{document}

\author{Tom Bachard \and Thomas Maugey}

\title{SMIC: semantic multi item compression based on CLIP dictionary}

\date{Supplemantary materials}

\maketitle

This supplementary document contains experiments and results that have not made to the main document due to the lack of space or because they do not quite fit in the flow of the main document.

\section*{III. Semantic linearity in the latent space}

\begin{figure}[H]
    \centering
    \includegraphics[width=\textwidth]{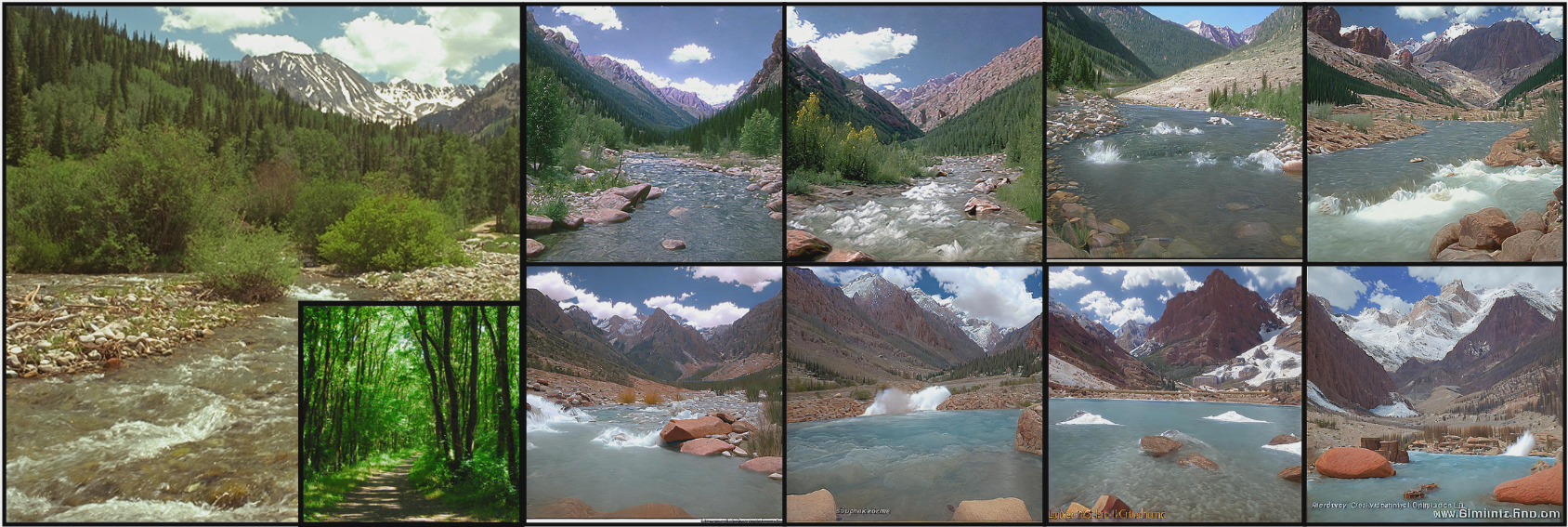}
    \caption{Another example of semantic subtraction. (Left) Input image and the semantic to subtract. (Right) Left to right: Images generated from $f(\bx_i)-\alpha f(\bx_2)$. Where $\alpha=i/8,\ i\in[1...8]$.}
    \label{fig:lin_sub}
\end{figure}

Fig.~\ref{fig:lin_sub} presents another example of semantic subtraction from the same input. In this experiment, the forest part of the input is progressively removed as $\alpha$ increases.

\begin{figure}[H]
    \centering
    \includegraphics[width=\textwidth]{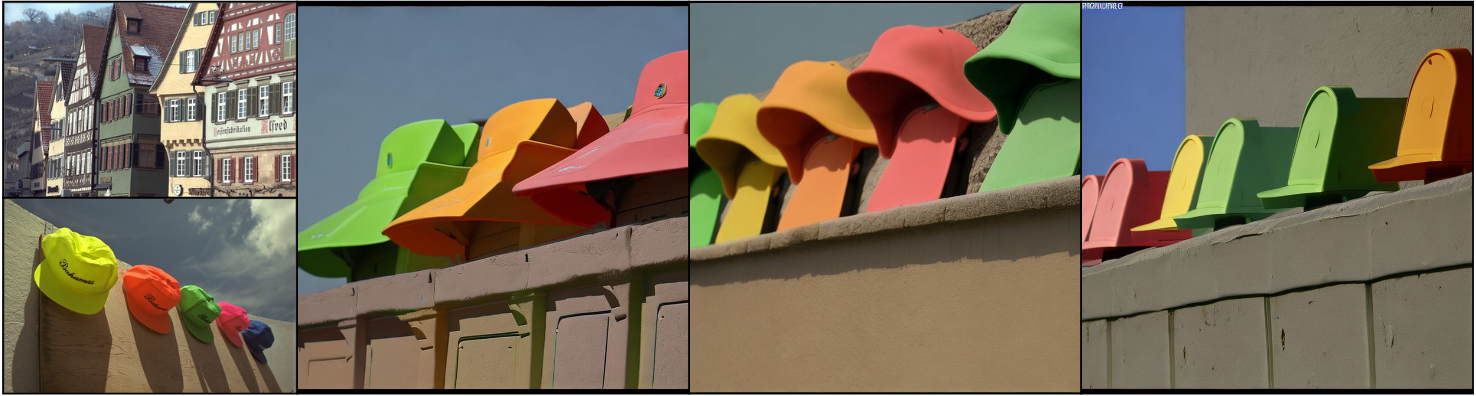}
    \caption{Generated images from semantic addition, $\alpha=0.5$. (Left) Input images. (Middle to Right) Generated images from the same latent vector.}
    \label{fig:gen_rand}
\end{figure}

We observe that generating images from the semantic addition of two images that does not make sense in the real-world results in images of lesser quality, as the generated images would depict non-sens.

\begin{figure}[H]
    \centering
    \includegraphics[width=\textwidth]{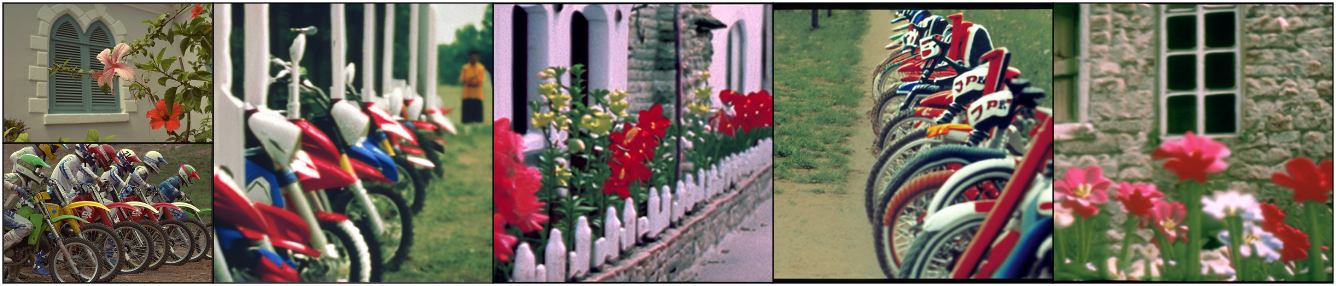}
    \caption{Generated images from semantic addition, $\alpha=0.5$. (Left) Input images. (Middle to Right) Generated images from the same latent vector.}
    \label{fig:gen_fifty}
\end{figure}

Another problem encountered while exploring semantic additions is the fact that for some example, the semantics do not add in the result image. Indeed, from the same latent vector, the generated images either converge toward the first image's semantic, or towards the second's, but never in between. We assume that these limitations come from the training method of both the encoder and the generator, as it cannot make sense of concepts it has not encountered.

\section*{IV. Learning a semantic dictionary from an image database.}

\subsection*{C. Interpretation of atom's semantic}

\begin{figure*}[htbp]
    \centering
    \includegraphics[width=\textwidth]{figures/semantic dict/atoms_32.png}\vspace{.3cm}
    \includegraphics[width=\textwidth]{figures/semantic dict/atoms_64.png}\vspace{.3cm}
    \includegraphics[width=\textwidth]{figures/semantic dict/atoms_128.png}
    \caption{Images generated from a dictionary learn on Landscape for $\lambda=0.5$ and various values of $n_a$. (Top to bottom) Generated images for $n_a=32,\ 64,\ \text{and } 128$ respectively.}
    \label{fig:atoms_ex}
\end{figure*}
    
Fig.~\ref{fig:atoms_ex} shows images generated from dictionaries learned on the Landscape dataset. For this experiment, we settled with $\lambda=0.5$ and $n_a\in[2^6, 2^7, 2^8]$. First, we observe that UnCLIP is able to generate coherent images from each latent (with one exception for the $8^{\text{th}}$ atom with $n_a=64$) with clear defined semantic in each. The second observation concerns the semantic of the images. Indeed, the image collection from which this dictionary has been learned on is Landscape; \emph{i.e.,} a dataset composed of images of sceneries (mountains, beaches, forests, etc.), and those characteristics are presented in the top atoms of each dictionary. More specifically, Landscape is made of seven classes, one of them being “landscapes Japan”. We then always observe a Japanese-like temple in the fourth position of each dictionary. For $n_a=64$ we can also observe a view of Mount Fuji, present in the initial data collection. Lastly, we observe that the high-level description of each atom is, for the most part, coherent from a value of $n_a$ to another. Indeed, we always have a mountain view with a grand valley in first position, a beach in second position, an Japanese-like temple in fourth, a field in fifth, a rocky waterfall in ninth, etc. From this observation, we can hypothesize that regardless of the size of the dictionary, the first atoms always represent more or less the same broad semantic of the data collection. Then, the following atoms focus on more and more specific semantic concepts from the database. We conclude from this experiment that the dictionaries learned from the image collection are of a semantic nature. 

\subsection*{D. Reconstructing images with a semantic dictionary}

\begin{figure}[H]
    \centering
    \includegraphics[width=\textwidth]{figures/semantic dict/decomp2.png}\vspace{.6cm}
    \includegraphics[width=\textwidth]{figures/semantic dict/decomp3.png}\vspace{.6cm}
    \includegraphics[width=\textwidth]{figures/semantic dict/decomp4.png}\vspace{.6cm}
    \caption{More examples of decomposition in learned dictionary with $\alpha =0.75$ and $n_a=32$. (Left) Input image. (Right) Non-null atom and their associated coefficient.}
    \label{fig:decomp}
\end{figure}

Fig.~\ref{fig:decomp} show more examples for the semantic decomposition of images in the dictionary, and their associated coefficients.

\subsection*{E. Image representation with a dictionary}

\begin{figure}[H]
    \centering
    \includegraphics[width=\textwidth]{figures/semantic dict/recon_resi_1.png}
    \includegraphics[width=\textwidth]{figures/semantic dict/recon_resi_3.png}
    \caption{Image generated from the projection and the residual of an image outside the data collection used to learn the dictionary. (Left) Input image. (Middle) Image generated via $\hat{\bz}$. (Right) Image generated via $\bar{\bz}$.}
    \label{fig:proj_resi}
\end{figure}

Fig.~\ref{fig:proj_resi} shows more examples of the dictionary projection generation and the residual generation from an image originally outside the dataset. We clearly observe the semantic separation of the origin database and the semantic that is outside of it.

\section*{V.Semantic multi-item compression.}

\subsection*{C. Rate Semantic Fidelity Optimization}

\textbf{Number of atoms in the dictionary:}

\begin{figure}[H]
    \centering
    \includegraphics[width=0.32\textwidth]{figures/MIC - expe/CC_atom.png}
    \hfill
    \includegraphics[width=0.32\textwidth]{figures/MIC - expe/CSS_atom.png}
    \hfill
    \includegraphics[width=0.32\textwidth]{figures/MIC - expe/BSS_atom.png}
    \caption{Evolution of semantic metrics regarding the number of atoms in the semantic dictionary. (Left to right) CC, CSS, BSS.}
    \label{fig:atom_eval}
\end{figure}

Fig.\ref{fig:atom_eval} shows experimental validation for the evolution of the different metrics regarding the evolution of the number of atoms in the dictionary.

\textbf{Sparsity of the coefficients:}

    \begin{figure}[H]
        \centering
        \includegraphics[width=0.32\textwidth]{figures/MIC - expe/CC_spar.png}
        \hfill
        \includegraphics[width=0.32\textwidth]{figures/MIC - expe/CSS_spar.png}
        \hfill
        \includegraphics[width=0.33\textwidth]{figures/MIC - expe/BSS_spar.png}
        \caption{Visual evolution of semantic metrics regarding the $\lambda$ in the sparse decomposition. (Left to right) CC, CSS, BSS.}
        \label{fig:spar_eval}
    \end{figure}

Fig.\ref{fig:spar_eval} presents the evolution of the conservation of semantics regarding $\lambda$.

    \begin{figure}[H]
        \centering
        \includegraphics[width=\textwidth]{figures/MIC - expe/ex_spar.png}
        \caption{Evolution of generated images for different values of $\lambda$. (Right) Input image. (Left) Generated images for $\lambda\in [0, 0.5, 1, 1.5, 2]$.}
        \label{fig:spar_ex}
    \end{figure}

Visual examples are proposed in Fig.\ref{fig:spar_ex}, where the decreasing semantic coherence between the inputs and the outputs is clearly deteriorating as $\lambda$ increases.

\textbf{Atoms' bits per dimension:}

    \begin{figure}[H]
        \centering
        \includegraphics[width=0.32\textwidth]{figures/MIC - expe/CC_dict.png}
        \hfill
        \includegraphics[width=0.32\textwidth]{figures/MIC - expe/CSS_dict.png}
        \hfill
        \includegraphics[width=0.32\textwidth]{figures/MIC - expe/BSS_dict.png}
        \caption{Evolution of semantic metrics regarding the number of bits per dimension for the atoms in the semantic dictionary. (Left to right) CC, CSS, BSS.}
        \label{fig:b_dict_eval}
    \end{figure}

    \begin{figure}[H]
        \centering
        \includegraphics[width=\textwidth]{figures/MIC - expe/ex_b_dico.png}
        \caption{Visual evolution of generated images for different values of $b_{\text{dict}}$. (Right) Input image. (Left) Generated images for $b_{\text{dict}}\in [2,4,8,16,32]$.}
        \label{fig:b_dict_ex}
    \end{figure}

Both the quantitative experiments, presented in Fig.~\ref{fig:b_dict_eval}, and the quantitative experiments, presented in Fig.\ref{fig:b_dict_ex}, confirm the observation made about Clip's latent space resistance to quantization.

\textbf{Coefficients' bitrate:}

    \begin{figure}[H]
        \centering
        \includegraphics[width=0.32\textwidth]{figures/MIC - expe/CC_coef.png}
        \hfill
        \includegraphics[width=0.32\textwidth]{figures/MIC - expe/CSS_coef.png}
        \hfill
        \includegraphics[width=0.32\textwidth]{figures/MIC - expe/BSS_coef.png}
        \caption{Evolution of semantic metrics regarding the number of atoms in the semantic dictionary. (Left to right) CC, CSS, BSS.}
        \label{fig:b_coef_eval}
    \end{figure}

    \begin{figure}[H]
        \centering
        \includegraphics[width=\textwidth]{figures/MIC - expe/ex_b_coef.png}
        \caption{Visual evolution of generated images for different values of $b_{\text{coef}}$. (Right) Input image. (Left) Generated images for $b_{\text{coef}}\in [2,4,8,16]$.}
        \label{fig:b_coef_ex}
    \end{figure}

In the same vein as $b_{\text{dict}}$, the impact of $b_{\text{coef}}$ is minimal, up to a certain threshold, as depicted in Fig.\ref{fig:b_coef_eval}. This indicates that we can propose a harsh quantization on the coefficients, while still being able to conserve the semantic of the inputs. A qualitative evaluation is proposed in Fig.~\ref{fig:b_coef_ex}.

\subsection*{D. Comparison to the state-of-the-art}

    \begin{figure}[htbp]
        \centering
        \includegraphics[width=0.48\textwidth]{figures/MIC - expe/rdo_100_CC_convex_hull_param.png}
        \hfill
        \includegraphics[width=0.48\textwidth]{figures/MIC - expe/rdo_10000_CC_convex_hull_param.png}
        \caption{Rate-distortion optimization curves for CC and the associated $(n_a, b_{\text{coef}}, b_{\text{dico}}, \lambda)$ parameter sets. (Left) $n=100$. (Right) $n=10000$}
        \label{fig:rdo}
    \end{figure}

\begin{figure}[htpb]
    \centering
    \includegraphics[width=\linewidth]{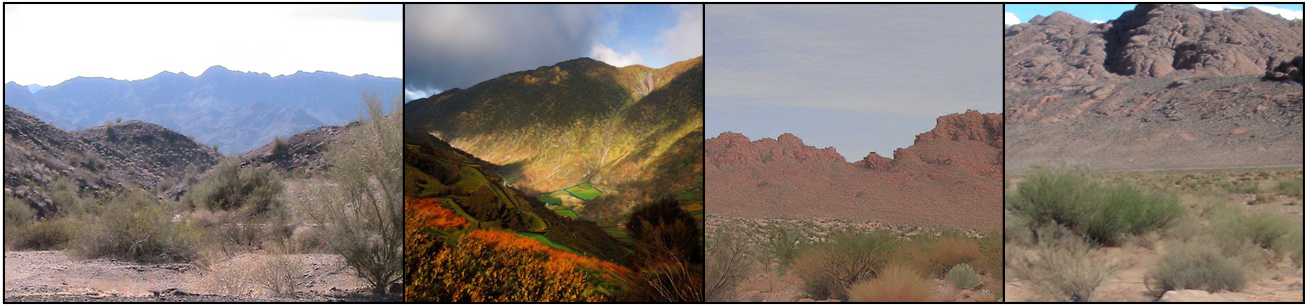}
    \caption{Images generated from our different models. (Left to right) Input image. Image generated via the \emph{low} model, via the \emph{medium} model, and via the \emph{high} model}
    \label{fig:model_ex_2}
\end{figure}

\begin{figure}[htpb]
    \centering
    \includegraphics[width=\linewidth]{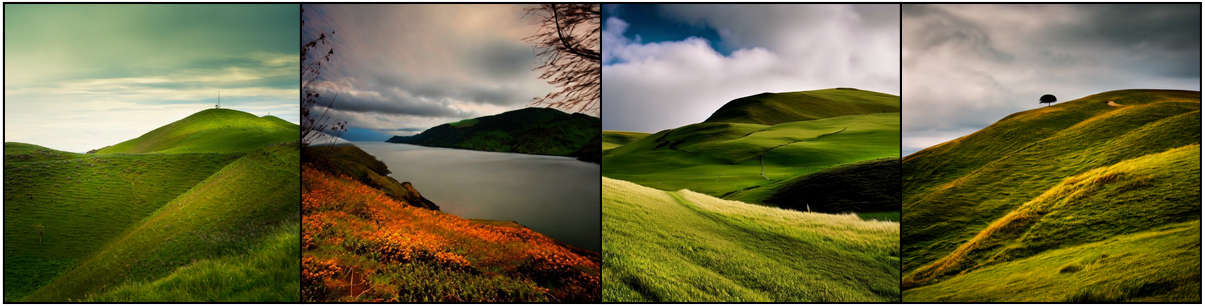}
    \caption{Images generated from our different models. (Left to right) Input image. Image generated via the \emph{low} model, via the \emph{medium} model, and via the \emph{high} model}
    \label{fig:model_ex_3}
\end{figure}

Images generated with the different codecs proposed in Sec. V.D. We observe that the images generated from the \emph{low} have lost import semantic details during the compression process. This artifact is minimal with the \emph{medium} model, and absent with the \emph{high} model.